\documentclass[prd,reprint,superscriptaddress]{revtex4-1}
\usepackage{amsmath}
\usepackage{amssymb}
\usepackage{graphicx}
\usepackage[caption=false]{subfig}
\usepackage{color}
\usepackage[percent]{overpic}
\usepackage{bm}
\usepackage{rotating}
\usepackage{hyperref}
\usepackage{pbox}
\usepackage{array}
\usepackage{physics}
\usepackage{mathtools}
\usepackage{enumerate}
\usepackage{leftidx}

\newcommand{\ii}{\mathrm{i}}

\usepackage{suffix}
\DeclarePairedDelimiterX\MeijerM[3]{\lparen}{\rparen}%
{\begin{smallmatrix}#1 \\ #2\end{smallmatrix}\delimsize\vert\,#3}

\newcommand\MeijerG[8][]{%
	G^{\,#2,#3}_{#4,#5}\MeijerM[#1]{#6}{#7}{#8}}

\WithSuffix\newcommand\MeijerG*[7]{%
	G^{\,#1,#2}_{#3,#4}\MeijerM*{#5}{#6}{#7}}

\allowdisplaybreaks

\usepackage[export]{adjustbox}

\begin{document}
	
	\title{Relativity and quantum optics: accelerated  atoms in optical cavities}

	\author{Richard Lopp}
	\affiliation{Department of Applied Mathematics, University of Waterloo, Waterloo, Ontario, N2L 3G1, Canada}
	\affiliation{Institute for Quantum Computing, University of Waterloo, Waterloo, Ontario, N2L 3G1, Canada}

	\author{Eduardo Mart\'{i}n-Mart\'{i}nez}
	\affiliation{Department of Applied Mathematics, University of Waterloo, Waterloo, Ontario, N2L 3G1, Canada}
	\affiliation{Institute for Quantum Computing, University of Waterloo, Waterloo, Ontario, N2L 3G1, Canada}
	\affiliation{Perimeter Institute for Theoretical Physics, 31 Caroline St N, Waterloo, Ontario, N2L 2Y5, Canada}
	
		\author{Don N. Page}
	\affiliation{Theoretical Physics Institute, Department of Physics, 4-183 CCIS, University of Alberta, Edmonton, Alberta T6G 2E1, Canada}
	
	
		\begin{abstract}





We analyze the physics of accelerated particle detectors  (such as atoms) crossing optical cavities. In particular we focus on the detector response as well as on the energy signature that the detectors imprint in the cavities. In doing so, we examine to what extent the usual approximations made in quantum optics in cavities (such as the  single-mode approximation, or the dimensional reduction of 3+1D cavities to simplified 1+1D setups) are acceptable when the atoms move in relativistic trajectories. We also study the dependence of these approximations on the state of the atoms and the relativistic nature of the trajectory. We find that, on very general grounds, and already in the weak coupling limit, single- and few-mode approximations, as well as 1+1D dimensional reductions, yield incorrect results when relativistic scenarios are considered.

	
		
		\end{abstract}
	
	\maketitle
	\section{Introduction}

To study the physics of atoms inside optical cavities sometimes approximations coming from quantum optical considerations are employed. For example, it is common to carry out the \textit{single-mode approximation} (or perhaps in some cases a few-mode approximation) where the number of modes in the cavity is reduced to a subset of close-to-resonance modes that the atom interacts with. Another common approximation is to consider 1+1D cavities neglecting the fact that the cavities are implemented in 3+1 dimensional spacetime. This last consideration may in principle seem reasonable in the case of, for instance, optical fibres that are very long as compared to their cross section.

 Although there are many treatments of quantum optics in a cavity in which the full spatial and temporal mode structures are considered, the number of cases where these approximations  are used is vast. For instance, among many, see \cite{scully1, scully2,steck2006,prants1999,miller2005,deb1996,sabin2017,doherty2000} for the single- or few-mode approximation, or, e.g., \cite{fuentes2011,aida2014,wang2014,kimble2012,liberato} for the usage of 1+1D cavities. While simplifying the problem, sometimes the rationale for these simplifications remains to be justified, above all in relativistic regimes. In a similar spirit, in \cite{liberato} the authors investigate in 1+1D the validity of the single-mode approximation inside a cavity with a stationary qubit, but still within a 1+1D framework and limited only to the ultra-strong coupling regime. 

In this paper we will analyze  whether the common approximations of quantum optics are valid in the weak coupling limit in a 3+1D cavity setup for a moving two-level particle detector like an atom. That comprises analyzing the soundness of the single- and few-mode approximation and of the approximation consisting of reducing the 3+1D model to a 1+1D problem for long cavities of small cross section. 

This is particularly relevant in the context of the Unruh effect \cite{fulling,davies,unruh1} within a cavity, i.e. for relativistic trajectories of particle detectors in cavities. Specifically, this is of importance in the light of relatively recent proposals for experiments for the detection of the Unruh effect involving optical cavities \cite{scully1,scully2,shresta,peres, donpage2017arxiv}.

We will characterize when and how any of those approximations are acceptable  in the ideal case where the optical cavities are considered to be made out of perfect conducting plates. We will see that neither the few-mode approximation nor the consideration of 1+1 dimensional cavities to approximate long (optical fibre-like) 3+1D cavities are generally justified for moving atoms in cavities when relativistic trajectories are considered (such as those commensurate with the Unruh effect). 


	
	\section{Setup}
	\subsection{Objectives}
\begin{figure}[!h]
		\includegraphics[scale=1]{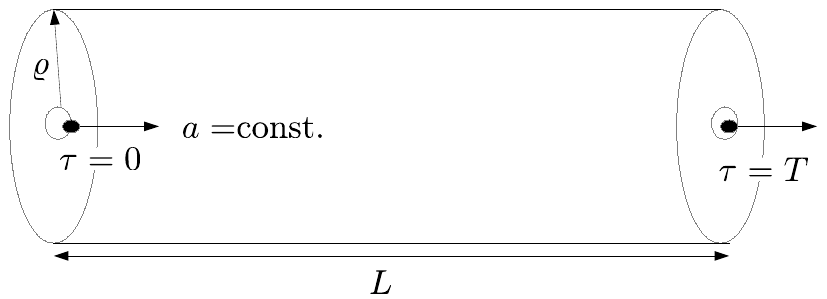}
		\caption{Detector moving with constant proper acceleration $a$ along the central axis of  a cylindrical cavity of length $L$.}
		\label{fig}
	\end{figure}

	We consider an accelerated particle detector inside a cylindrical cavity. We wish to analyze the effects of the detector on the quantum field inside the cavity due to its acceleration, as well as the detector's response. The scenario is depicted in Fig.~\ref{fig}. The first thing we will analyze is in which field modes the energy deposited after the detector crossed the cavity depending on the detector's initial state and its trajectory. In other words, which field modes will get excited and how does the energy distribute in the field modes after the passing of the detector through the cavity after one run.

	We will model the detector as an Unruh-DeWitt particle detector consisting of two energy levels separated by a gap $\Omega$, and the field will be taken to be a massless scalar field in $3+1$ dimensions. This setup captures all the relevant features of the light-matter interaction neglecting exchange of angular momentum between field and detector (for more details check \cite{PhysRevD.14.870, deWitt}, and  section II  of \cite{Pozas2016}). 

In addition to the analysis above, we will further compare the state of the field  after the cavity is crossed by an accelerated detector to the case where the detector is moving with constant velocity, and also comparing to a non-relativistic approximation to the trajectory of the detector. The former will show  whether the state of the field has any unique signature deriving from the acceleration as opposed to non-accelerated detectors.

Moreover, as a measure of validity of the single- or few-mode approximation, we will study the transition probabilities  after crossing the cavity and how much the predictions of the single- and few-mode approximation deviate from the exact results.

Finally, we will investigate whether  it is possible to use a 1+1D model to reproduce the 3+1D model where the length of the cavity is much larger than its radius, (i.e. an  \lq optical fibre' experiment), both in the case of a massless 1+1D model and in the refined case of a massive 1+1D model whose mass results from the effective reduced dimensions of spacetime. For this we will also compare the predictions for the atomic dynamics.


	\subsection{Time evolution}
	
Since we are considering a cylindrical cavity with Dirichlet boundary conditions modeling a perfectly reflecting plate, first the massless Klein-Gordon equation  is solved in cylindrical coordinates (see Appendix~\ref{derivation}), so that the quantized scalar field reads
	\begin{align}
	\hat{\phi}(r,\varphi,z,t)= \sum_{\substack{m=0 \\ n,l=1}}^{\infty}\left(	u_{m l n } \hat{a}_{m l n } + 	u^*_{m l n } \hat{a}^{\dagger}_{m l n }\right),
	\end{align}
	where the creation and annihilation operators $ \hat{a}^{\dagger}_{m l n }$ and $\hat{a}_{m l n }$ have the canonical commutation relations. $n$ is the longitudinal quantum number, and $m$ and $l$ are the quantum numbers corresponding to transversal degrees of freedom. The field modes $u_{m l n }$ have the form
	\begin{align}
	u_{m n l}(r,\varphi,z,t)&=A_{m  l n}  e^{\ii m \varphi}   e^{ -\ii \omega t } \sin(\frac{n \pi}{L} z) J_m\left(\frac{x_{m l}}{\varrho} r\right), \label{modes} \\
	A_{m l n}&= \frac{1}{\varrho \sqrt{L \pi \omega} J_{m+1}(x_{m l}) },\\
	\omega&=\sqrt{\frac{x^2_{m l}}{\varrho^2}+ \frac{n^2 \pi^2}{L^2}},\label{energ}	
	\end{align}
	with $x_{m l}$ being the $l$-th zero of the $m$-th Bessel function of the first kind $J_m$.

	After detector crossing the initial state $\hat \rho_{0}$ of joint system of detector and cavity transforms to
	\begin{align}
	\hat \rho_{\text{d},\phi} &=\hat U \hat \rho_0 \hat{U}^{\dagger}, 
	\end{align}
	where
	\begin{equation}
	\hat{U} =\mathcal{T} \exp( -\ii \int_{0}^{T} \text{d}\tau \hat{H}_I(\tau)).\label{evolution}
	\end{equation}
	$\mathcal{T}$ denotes the time-ordering operation, and the integration limits $0$ and $T$ correspond to the times at which the detector enters and exits the cavity in the detector's frame respectively (see Fig.~\ref{fig}). The Unruh-DeWitt interaction Hamiltonian in the interaction picture reads
	\begin{align}
	\hat H_I(\tau)=\lambda \hat \mu(\tau) \hat{\phi}(\bm x(\tau),t(\tau)), 
	\end{align}
	where $\lambda$ is the coupling strength between detector and field, and $\hat \mu(\tau)=e^{\ii \Omega \tau}\hat \sigma^+ + e^{-\ii \Omega \tau}\hat \sigma^-$ is the detector's monopole moment ($\hat \sigma_x$ in the interaction picture) with $\hat \sigma^{+/-}$ being the SU(2) ladder operators, and $\tau$ is the proper time of the detector. In the Hamiltonian we did not include a time-dependent switching function as inside the cavity the coupling strength between detector and field will be assumed to be constant. This is not a problem since the Dirichlet boundary conditions ensure that there are no UV divergences despite the finiteness of the interaction. 
	
	The time-evolved state will be calculated by a perturbative Dyson expansion of \eqref{evolution}, granted the relevant parameters are small enough:
	\begin{equation}
	\hat U=\openone\underbrace{-\ii\int_{0}^{T}\!\!\! \dd \tau\,\hat H_I(\tau)}_{\hat U^{(1)}}\underbrace{-\!\!\int_{0}^{T}\!\!\!\!\text{d}\tau\int_{0}^{\tau}\!\!\!\text{d}\tau^{\prime}\,\hat H_I(\tau)\hat H_I\left(\tau^{\prime}\right)}_{\hat U^{(2)}}+\dots\label{eq:evo}
	\end{equation}
	Thus, to second order in the coupling constant $\lambda$ the evolved state takes the form
	\begin{align}
	\hat \rho_{\text{d},\phi}&= \hat \rho_0 + \hat U^{(1)} \hat \rho_0 + \hat \rho_0 \hat U^{(1) \dagger} \nonumber\\
	&\quad +  \hat U^{(2)} \hat \rho_0 + \hat \rho_0 \hat U^{(2) \dagger} +  \hat U^{(1)} \hat \rho_0 \hat U^{(1) \dagger} + \mathcal{O}(\lambda^3).
	\end{align}
	We assume that the field is initially in the vacuum state, and that field and detector start out uncorrelated:
	\begin{align}
	    \hat\rho_0=\hat \rho_\text{d}\otimes \ketbra{0}{0}.
	\end{align}

	For the initial state of the detector we assume  it is either in the ground state $  \ket g$ or in the excited state $\ket e$. Then after interaction time $T$ between detector and field, the final state of the field reads to second order in both cases
	\begin{equation}
	\hat \rho_\phi=\ket 0\!\!\bra{0}  + \tr_{\text{d}}\left( \hat U^{(1)} \hat \rho_0 \hat U^{(1) \dagger}\right) + \left(\tr_{\text{d}}\left( \hat U^{(2)} \hat \rho_0\right) + \text{H.c.}\right). \label{field rho}
	\end{equation}
	Note that in \eqref{field rho} there are no first order terms as the detector starts in an energy eigenstate and therefore  $\tr_{\text{d}}\big( \hat U^{(1)} \hat \rho_0\big)=0$, thus $\mathcal{O}(\lambda^2)$ is the leading order.


\section{Validity of a single (or few) mode approximation}

We begin first assessing the (in)validity of the single-mode approximation in relativistic scenarios. We will do this in two ways:
\begin{enumerate}
 \item We will compute what modes become non-negligibly excited (what is the energy spectrum) in the field in two scenarios: first after an accelerated  detector crossed the cavity in the longitudinal direction, and secondly for a detector with constant velocity. 
\item We will calculate by how much the few-mode approximation fails to predict the transition probability of such detector.
\end{enumerate}

Then we will repeat the analysis with the detector following a constant velocity trajectory. As we will see, it is not true that most field excitations remain confined to near-resonant modes.

	\subsection{Energy spectrum in the field and transition probabilities for accelerated detectors}
	We consider a detector initially at rest at the entrance of the cavity $(r,\varphi,z,t)=0$, and its subsequent constant proper acceleration $a$ in the longitudinal direction $z$. The detector's worldline in the cavity frame parametrized by its proper time $\tau$ is
	\begin{align}
	z(\tau)=\frac{1}{a}\left(\cosh(a \tau)-1\right), ~t(\tau)=\frac{1}{a}\sinh(a \tau), ~r,\varphi=0.\label{a world}
	\end{align}
	Hence, the field's mode functions become
	\begin{align}
	u_{m l n}(\tau)= \delta_{m 0} A_{m l n}   e^{- \ii \frac{\omega}{a}\sinh(a \tau) } \sin(\frac{n \pi}{a L}\left(\cosh(a \tau)-1\right)\!), \label{longit}
	\end{align}
	where all contributions with $m\neq 0$ vanish since  for $r=0$ we have that \mbox{$J_m(0)=\delta_{m 0}$}. 
	The (detector's proper time) duration of the interaction inside the cavity is 
$	T= \text{arccosh}\!\left(a L +1\right)/a.$
	As can be seen in Appendix~\ref{derive longit},  the number expectation value of modes with quantum number $m=0$ is
	\begin{align}
	    N_{l, n}=\lambda^2   \left|\int_0^T  \dd \tau e^{\pm \ii  \Omega\tau } u_{0ln}^*(\tau) \right|^2,  \label{number}
	    \end{align}
	where $l, n >0$, and the $\pm$ is there to  indicate that for the $+$ sign the initial state of the detector is the ground state and the $-$ sign yields the result for the detector initially in the excited state. 
	We show as well in Appendix~\ref{derive longit} that, to leading order, $\sum_{l,n} N_{l,n}$ equals the probability of finding the detector in a different state than initially started after cavity crossing (again, this means that with a $+$ we start in $\ket{g}$ and we will get the vacuum excitation probability, and with a $-$ we start in $\ket{e}$ and get the probability of spontaneous emission). We denote the vacuum excitation probability  $\mathcal{P}^{g\to e}$  and the probability of spontaneous emission $\mathcal{P}^{e\to g}$.
	
	Hence, we can take as a measure of validity of the single- or few-mode approximation the ratio
	\begin{align}
		    \frac{N_{\text{res}}}{\sum_{n,l=1}^\infty N_{l,n}} =\frac{\mathcal{P}^{g\to e/ e\to g}_\text{res}}{\mathcal{P}^{g\to e/ e\to g}},
		\end{align}
	where the subscript `res' indicates the contribution of the resonant mode (single or few if they are close in energy). This ratio is easy to justify: it tells us the relative magnitude of the contribution of the resonant mode(s) with respect to the full calculation where the single mode approximation is not assumed.

	\begin{figure*}[p]
		\centering
\includegraphics[scale=1]{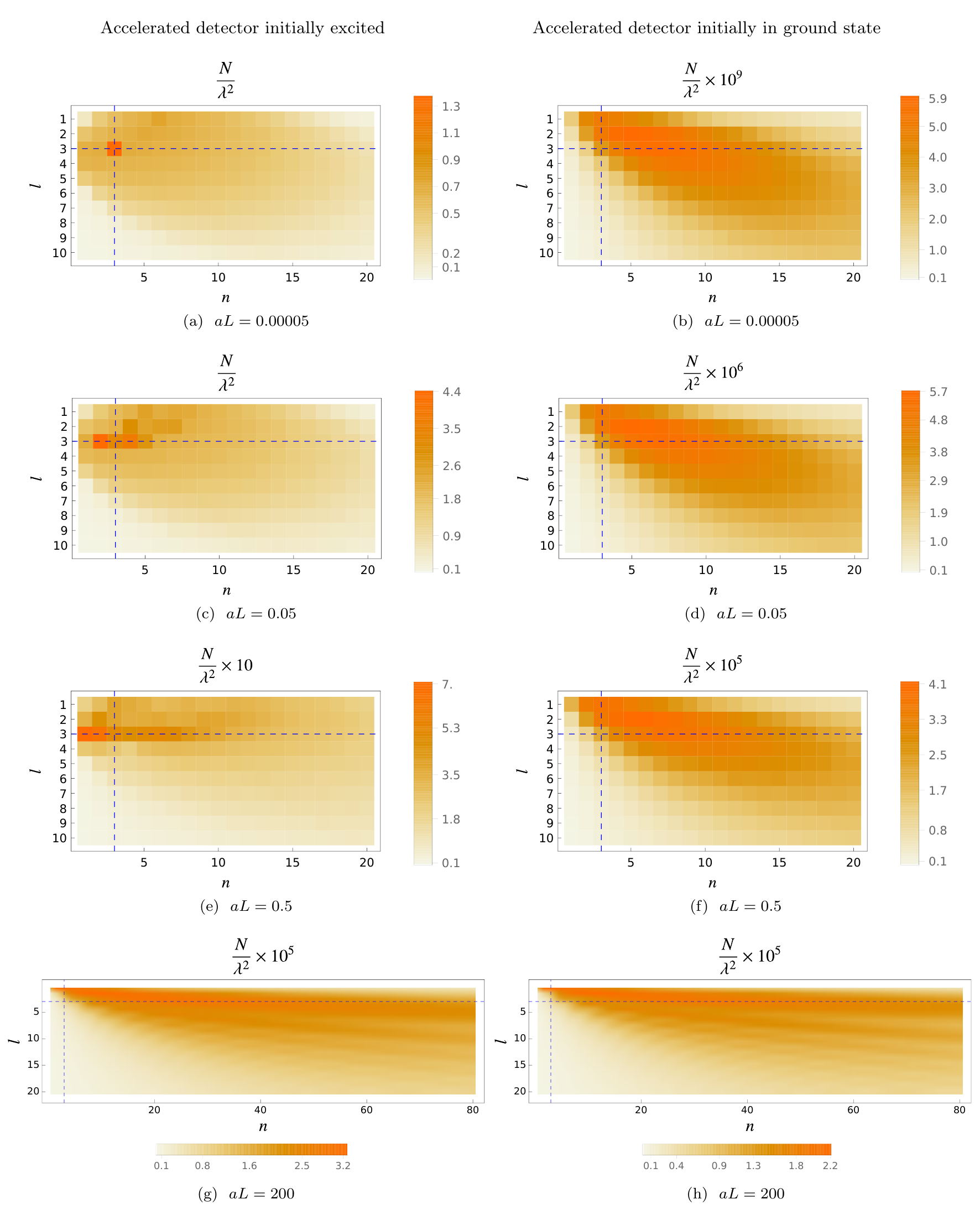}
		\caption{Number expectation value $N$ (proportional to the energy per mode) as a function of mode numbers $n$ and $l$ for an accelerated detector, comparing relativistic to non-relativistic behavior. Parameters are $\varrho/L=0.5$, $\Omega \varrho=10$, $\Omega L =20$ such that the detector's energy gap is resonant with $\omega$ for $(m, l, n)=(0, 3, 3)$ (intersection of dashed lines). (a, b)  $a L =0.00005$ (final velocity about $0.01$); 
    			(c, d) $a L =0.05$ (final velocity about $0.3$); (e, f)  $a L=0.5$  (final velocity about $0.75$); (g, h) $a L=200$ (final velocity of $0.99999$). With higher accelerations, the resonance of the excited case experiences a Doppler shift, which results in a broadening in the peak and a higher number of excited modes. }
		\label{compes}
	\end{figure*}
	
		\begin{figure*}[p]
		\centering
		
		Detector with constant velocity initially excited ~~~~~~Detector with constant velocity initially in ground state \\
\begin{tabular}[c]{cc}
				\subfloat[ $\bar{v}=0.005$]{\label{fig:right}%
				\includegraphics[scale=1,valign=t]{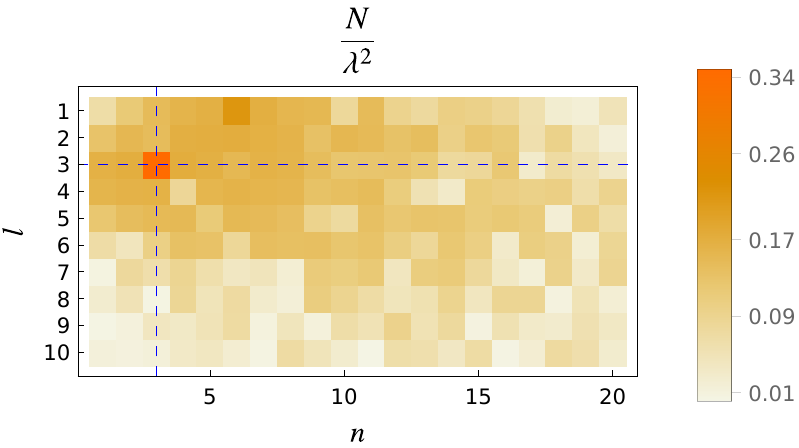}}
			& \subfloat[ $\bar{v}=0.005$ ]{\label{fig:left}%
			\includegraphics[scale=1,valign=t]{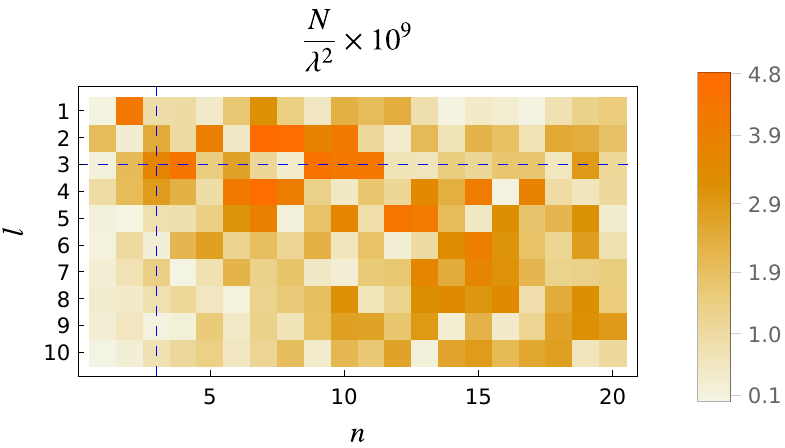}	}\\
			
				\subfloat[   $\bar{v}=0.16$]{\label{fig:right}%
				\includegraphics[scale=1,valign=t]{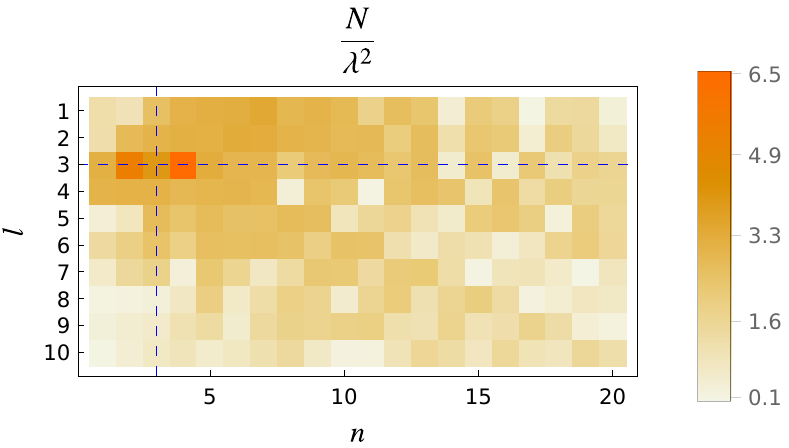}}
			& \subfloat[  $\bar{v}=0.16$  ]{\label{fig:left}%
			\includegraphics[scale=1,valign=t]{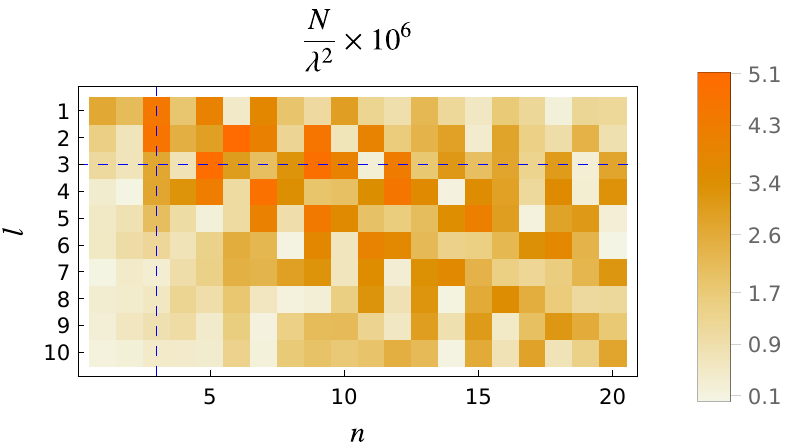}	}
			\\
			
				\subfloat[ $\bar{v}=0.45$]{\label{fig:right}%
				\includegraphics[scale=1,valign=t]{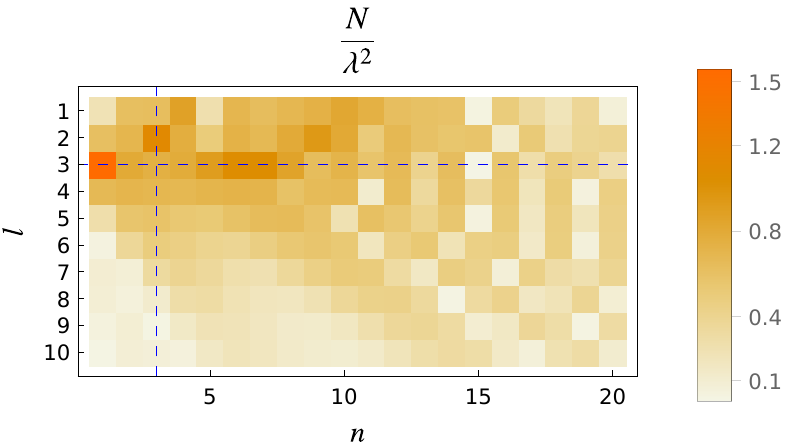}}
			& \subfloat[ $\bar{v}=0.45$  ]{\label{fig:left}%
			\includegraphics[scale=1,valign=t]{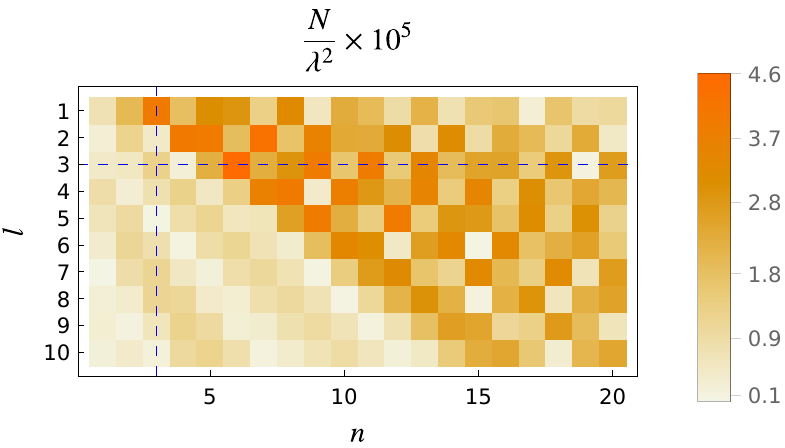}	}
			\\
			
				\subfloat[  $\bar{v}=0.995$]{\label{fig:right}%
				\includegraphics[scale=.61,valign=t]{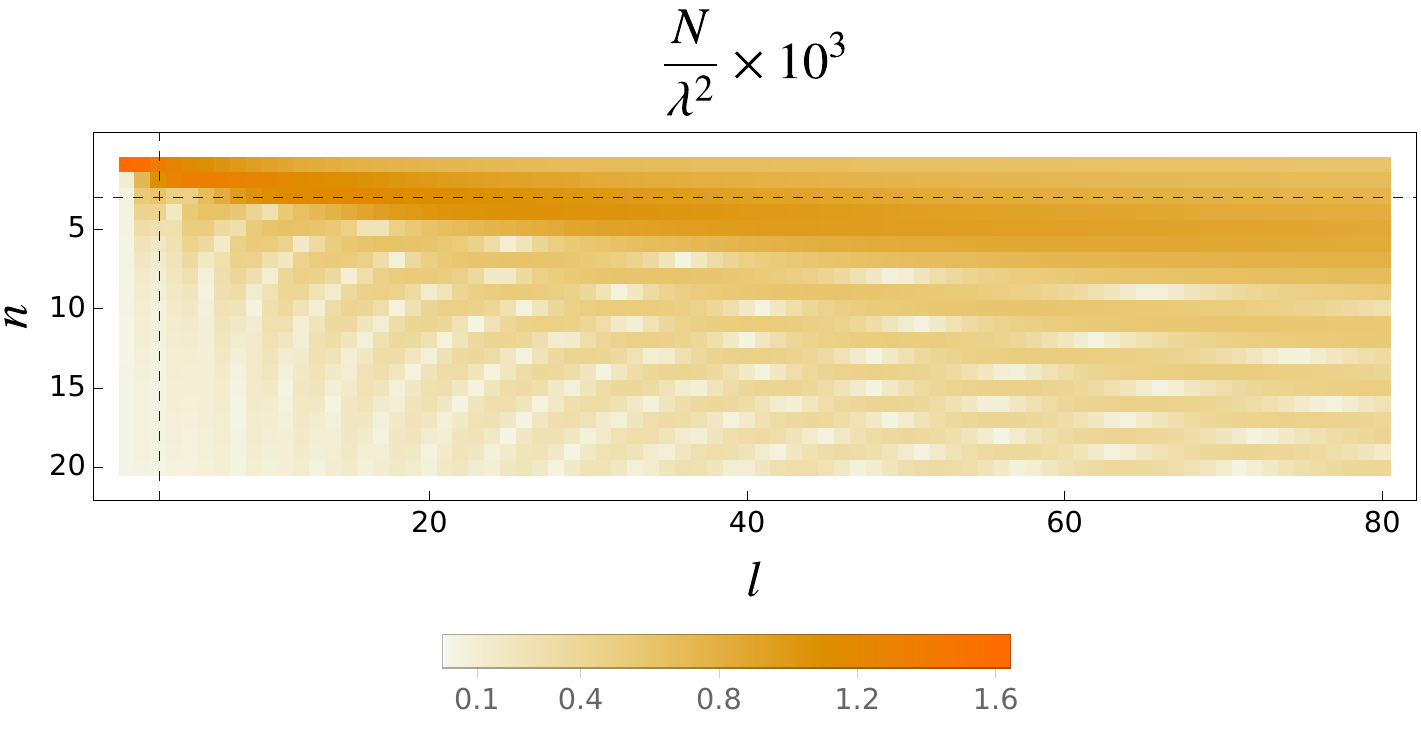}}
			& \subfloat[$\bar{v}=0.995$  ]{\label{fig:left}%
			\includegraphics[scale=.61,valign=t]{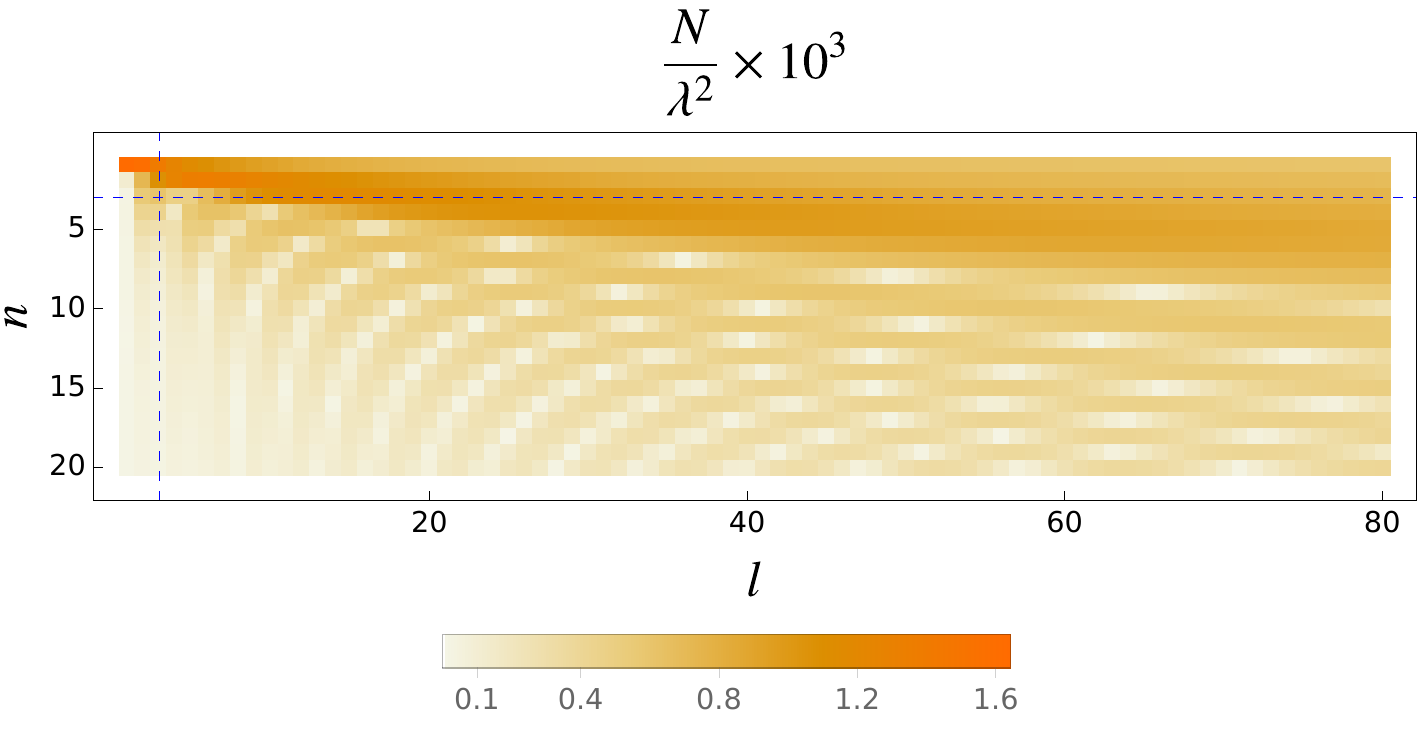}	}

		\end{tabular}
		\caption{Number expectation value $N$ of the cavity field as a function of mode numbers $n$ and $l$, comparing relativistic to non-relativistic behavior for the case of a detector with constant velocity.  Parameters are  $\varrho/ L=0.5$,  $\Omega \varrho=10$,  $\Omega L =20$ such that the detector's energy gap is resonant with $\omega$ for $(m, l, n)=(0, 3, 3)$ (intersection of dashed lines). (a, b) Velocity $\bar{v}=0.005$  (corresponding to $a L=0.00005$); 
			(c, d)  $\bar{v}=0.16$  (corresponding to $a L=0.05$);	(e, f) $\bar{v}=0.45$  (corresponding to $a L=0.5$); (g, h) $\bar{v}=0.995$  (corresponding to $a L=200$). The velocities are chosen such that the detector traverses the cavity in the same time interval (as seen form the cavity frame) as for the corresponding cases of an accelerated detector.}
		\label{velo1}
	\end{figure*}

				In Fig.~\ref{compes} the distribution of excitations  in the field modes is displayed for different accelerations for both detector settings. As can be seen clearly, for non-relativistic set-ups ($a L \ll 1$) in the case of the detector initially in the excited state  the excitations peak in the vicinity of the resonant frequency $\omega \approx \Omega$. However, for larger accelerations, and hence larger final velocities, excitations can be found far away from the resonance due to the relativistic Doppler effect.  If the detector starts out in the ground state, there is no peak around the resonant frequencies at all. This is due to the absence of a resonance in this regime as well as the fact that counter-rotating contributions become strong as the motion becomes more and more relativistic, and the detector spends less time to cross the cavity. Moreover, the initial ground state configuration  evolves to number expectation values which are several orders of magnitude less than for the case where the detector is initially excited. 
				
				In Table~\ref{ratio1} we present upper bounds for the transition probabilities.
			    We define resonant modes as such modes that are  within 2\%  of the detector's gap $\Omega$.
            	The table shows that for an excited detector in non-relativistic regimes the non-resonant contribution may be negligible, depending on the set of parameters. Nonetheless, going to relativistic accelerations will significantly increase the non-resonant contribution, as was expected from the Doppler shift, and also adding a mode spread of the energy deposited in the field. In principle, this renders the single-mode or few-mode approximation invalid in the excited case for high accelerations. If the detector enters the cavity in its ground state, things look even worse for the single-mode approximation: the resonant contribution is negligible for all regimes, and thus we cannot expect a mode few-approximation to be justified. 
            	
            	We have also performed a further study of the validity of the approximation depending on the different parameters of the problem, and it can be see in Appendix~\ref{parameterspace}.
            	
				

				\renewcommand{\arraystretch}{1.2}
				    \begin{table*}[ht]
				        \centering
				        \begin{tabular}{|c|| c|c|c|c|c|c| }
 				           \hline 
 				            Parameters  & \multicolumn{6}{c|}{$\varrho/L=1/2$, $\Omega L=5.75$, resonant with $(l,n)=(1,1)$} \tabularnewline \hline 
 				            $a L$ & $5\times 10^{-5}$&$5\times 10^{-4}$&$5\times 10^{-3}$&$5\times 10^{-2}$&$5\times 10^{-1}$&$200$\\ \hline 
				           $ \mathcal{P}_{\text{res}}^{e\to g}/ \mathcal{P}^{e\to g} \leq$ &1.00&1.00&1.00&0.97&0.48&$7\times 10^{-5}$ \\\hline
				            $ \mathcal{P}_{\text{res}}^{g\to e}/ \mathcal{P}^{g\to e} \leq$ & $6.6\times 10^{-3}$ &$6.5\times 10^{-3}$&$6.5\times 10^{-3}$&$6.4\times 10^{-3}$&$5.2\times 10^{-3}$&$6.8\times 10^{-5}$ \\ \hline
 				            \hline 
 				           Parameters  & \multicolumn{6}{c|}{$\varrho/L=1/2$, $\Omega L=20$, resonant with $(l,n)=(3,3)$} \tabularnewline \hline 
 				            $a L$ & $5\times 10^{-5}$&$5\times 10^{-4}$&$5\times 10^{-3}$&$5\times 10^{-2}$&$5\times 10^{-1}$&$200$\\ \hline 
				           $ \mathcal{P}_{\text{res}}^{e\to g}/ \mathcal{P}^{e\to g} \leq$ & 0.89&0.8 & 0.42 &0.13&0.06 & $3.3\times 10^{-5}$\\\hline
				           $ \mathcal{P}_{\text{res}}^{g\to e}/ \mathcal{P}^{g\to e} \leq$ & $6.2\times 10^{-4}$&$6.2\times 10^{-4}$&$6.2\times 10^{-4}$&$6.2\times 10^{-4}$&$5.4\times 10^{-4}$&$2.9\times 10^{-5}$\\ \hline
 				            \hline 
 				            Parameters  & \multicolumn{6}{c|}{$\varrho/L=1/2$, $\Omega L=50$, 10 resonant modes} \tabularnewline \hline 
 				            $a L$ & $5\times 10^{-5}$&$5\times 10^{-4}$&$5\times 10^{-3}$&$5\times 10^{-2}$&$5\times 10^{-1}$&200\\ \hline 
				          $ \mathcal{P}_{\text{res}}^{e\to g}/ \mathcal{P}^{e\to g} \leq$ &0.99&0.98&0.53&0.09&0.06&$5.5\times 10^{-4}$\\\hline
				            $ \mathcal{P}_{\text{res}}^{g\to e}/ \mathcal{P}^{g\to e} \leq$ & $1.5\times 10^{-3}$&$1.5\times 10^{-3}$&$1.5\times 10^{-3}$&$1.5\times 10^{-3}$&$1.4\times 10^{-3}$&$5.4\times 10^{-4}$ \\ \hline
 				           \end{tabular}
				        \caption{3+1D: Determining the validity of the single- and few-mode approximation by finding an upper bound to the ratio of the resonant contribution to the total spontaneous emission probability $\mathcal{P}^{e\to g}$ and vacuum excitation probability $\mathcal{P}^{g\to e}$, respectively. The resonant modes have been chosen such that they differ at most 2~\% energetically from the detector gap $\Omega$.
				        We have taken as the cut-offs for the sums over $n$ and $l$ $10^4$ and $200$, respectively.
				        }
				        \label{ratio1}
				    \end{table*}

				\subsection{Single-mode approximation for constant-velocity motion}
				We compare now to the case of constant velocity $\bar{v}$ ($a=0$) in order to clarify which signatures are due to acceleration and which are a mere artifact of the velocity. To that end, we choose as the worldline 
				\begin{align}
				z(\tau)=\gamma \bar{v} \tau, ~t(\tau)=\gamma \tau,~ r,\varphi=0.
				\end{align}
				We find the mode functions to be
				\begin{align}
				u_{m l n}(\tau)= \delta_{m 0} A_{m l n}   e^{- \ii \omega \gamma \tau} \sin(\frac{n \pi}{ L}\gamma \bar{v} \tau),
				\end{align}
				where again all contributions with $m\neq 0$ vanish. 
				The length of interaction as given by the proper time is 
			$	T'=L/\gamma \bar{v}$, 
				and we choose the velocity such that the detector will require the same time (with respect to the cavity frame) as the uniformly accelerated one  to traverse the cavity:
				\begin{align}
				\bar{v}=\left(1+\frac{2}{a L}\right)^{-1/2}.
				\end{align}
			    This allows to analytically solve the integral for the number expectation values (see \eqref{number}):
				\begin{align}
				N_{l, n}=& \lambda^2  \frac{2 \pi ( n  \bar{v})^2}{\omega L^3 (\varrho \gamma	J_1(x_{0 l}))^2 } \frac{1+(-1)^{n+1} \cos(\left(\omega \pm\frac{\Omega}{\gamma} \right) \frac{L}{\bar v})}{  \left[\left(\omega\pm\frac{\Omega}{\gamma}\right)^2-\left(\frac{n \pi \bar v}{L}\right)^2\right]^2}  ,\label{constv}
				\end{align} where again the top sign denotes the initial ground state and the bottom sign the initial excited state of the detector.
	Interestingly, $N_{l, n}$ can be zero due to the oscillatory behavior if 
\begin{align}
  \left(\omega\pm\frac{\Omega}{\gamma}\right) \frac{L}{\bar{v}}= \pi m,
\end{align}
given that $\omega\pm\Omega/\gamma \neq 0$ and $m\in \mathbb{Z}\!\setminus\!\{n\}$, or for all odd $n$ if  $\omega\pm\Omega/\gamma = 0$, i.e.\ if a corresponding (Lorentz transformed) field mode is exactly resonant with the detector's gap and the detector is initially excited. This is a manifestation of the phenomenon called `mode invisibility', that was introduced in \cite{modeinv1} and has been also used in quantum optics for non-demolition measurements \cite{modeinv2,modeinv3}.

%
%
%
%
					In Fig.~\ref{velo1} we show the results for constant velocity. For both detector initializations (ground and excited), the distribution is clearly distinguishable from the constant acceleration setting. Considering constant velocity, the distribution of number expectation values is sensitive to the velocity, having zeros as discussed before. However, an initially excited detector causes a relativistic Doppler broadening of the initial resonance. As in the previous case, the initial ground state detector does not exhibit a resonance in the distribution, rendering any single-mode approximation invalid for any regime.

\begin{figure}
				        \centering
				        \vspace{.2cm}
				       \hspace{-.5cm} \includegraphics[scale=0.52]{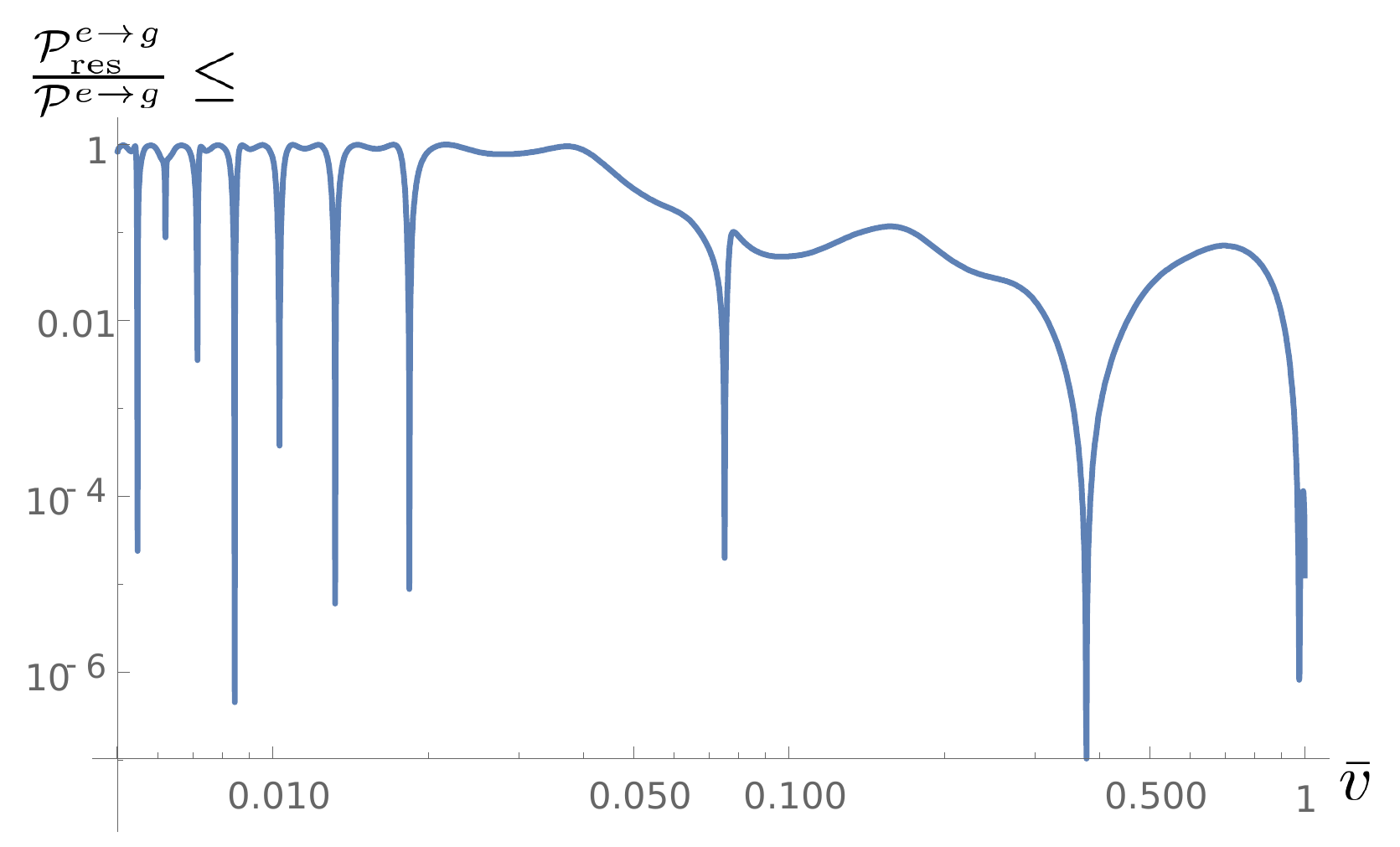}
				        \caption{Estimating an upper bound to the ratio of the resonant contribution to the spontaneous emission probability with $\varrho/L=0.5$, $\Omega L=20$ for an excited detector with constant velocity $\bar{v}$. The detector's energy gap is resonant with $\omega$ for $(m, l, n)=(0, 3, 3)$. We have chosen as the cut-offs for the sums over $n$ and $l$ $2000$ and $100$, respectively, and used a double-logarithmic plot.}
				        \label{ratio2}
				    \end{figure}

					In Fig.~\ref{ratio2} one can see  an upper bound to the ratio of the resonant contribution to the spontaneous emission probability, given an exemplary parameter setting. Here again, as was in the case for a uniformly accelerated detector, the single-mode approximation does not reproduce the distribution to a good fidelity for relativistic trajectories, and even for low velocities caution is required due to its oscillating behavior.


\section{Validity of a non-relativistic approximation}

	Another approximation that we will assess is the consideration that the trajectory of the detector undergoes non-relativistic motion and the trajectory is approximated by a Galilean constant accelerated motion. This greatly simplifies the mathematical treatment of the dynamics  and may be of interest for numerical reasons, but, as we will see, it is not enough that the final speed of the detector is non-relativistic to carry out this approximation: this approximation also fails to assess the amount of energy deposited in high enough energy modes regardless of the final speed of the detector.  We will see that only for a restricted number of modes the relative error in the mode numbers is small for a given acceleration.
	
				\begin{figure*}[!ht]
				 \centering
				 \textbf{Validity of non-relativistic approximation}\par\medskip
				Detector initially in excited state \\
				\begin{tabular}[c]{cc}
				\begin{tabular}[b]{c}
				\subfloat[  $a L=0.00005$   ]{\label{fig:right}%
					\hspace*{-0.4cm}	\includegraphics[scale=.87,valign=b]{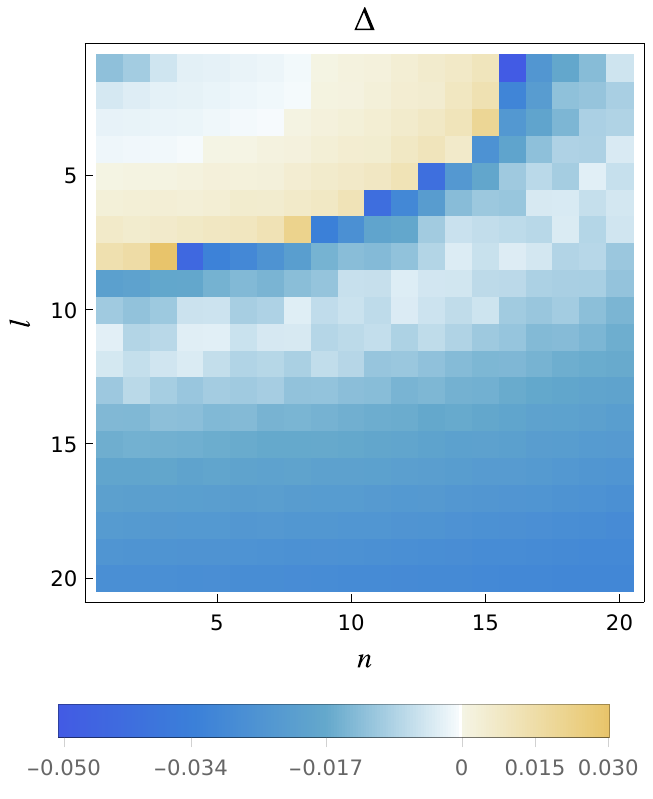}}
				
				\end{tabular}
				&
				\subfloat[ $a L=0.005$  ]{\label{fig:left}%

					\begin{tabular}[b]{c}
					\includegraphics[scale=.87,valign=b]{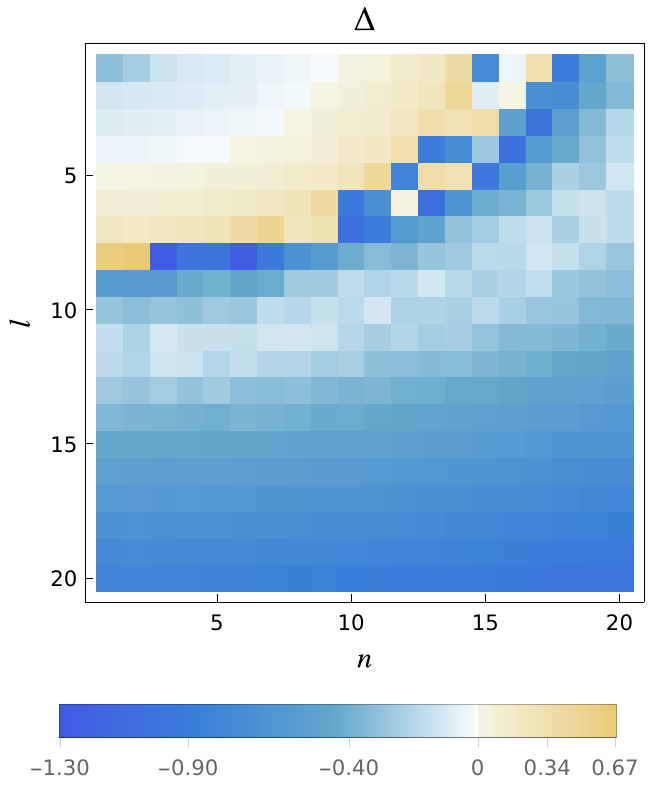}
					\end{tabular}
				}
				
				\subfloat[  $a L=0.05$ ]{\label{fig:left}%

					\begin{tabular}[b]{c}
					\includegraphics[scale=.87,valign=b]{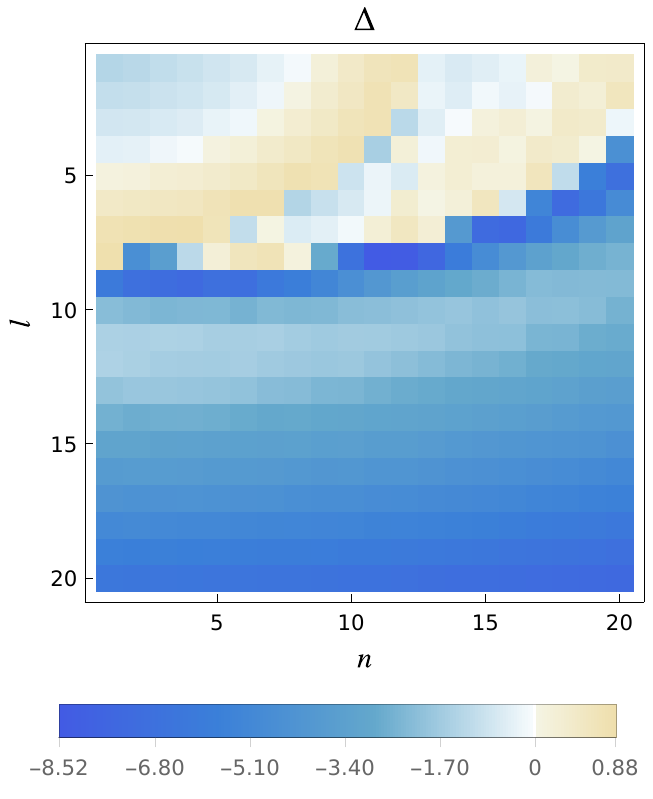}
					\end{tabular}
				}

				\end{tabular}
				 \centering
				 
				 \vspace{.5cm}
				 Detector initially in ground state \\
				\begin{tabular}[c]{cc}
				\begin{tabular}[b]{c}
				\subfloat[  $a L=0.00005$   ]{\label{fig:right}%
					\hspace*{-0.4cm}	\includegraphics[scale=.87,valign=b]{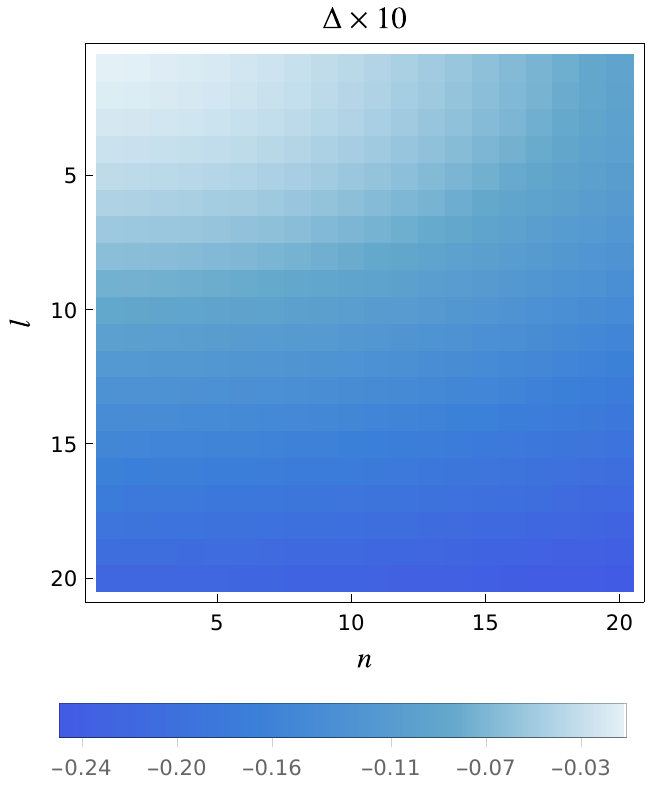}}
				
				\end{tabular}
				&
				\subfloat[ $a L=0.005$  ]{\label{fig:left}%

					\begin{tabular}[b]{c}
					\includegraphics[scale=.87,valign=b]{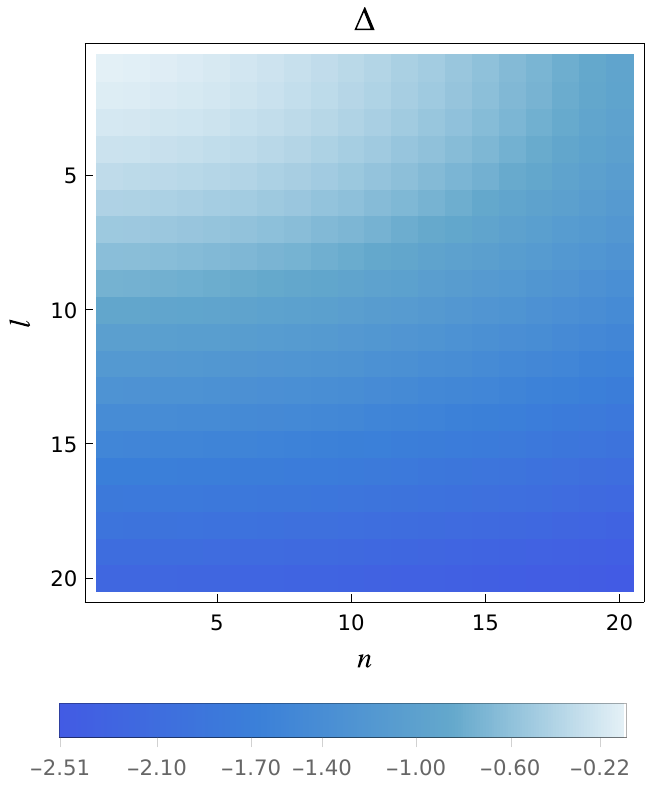}
					\end{tabular}
				}
				
				\subfloat[  $a L=0.05$ ]{\label{fig:left}%

					\begin{tabular}[b]{c}
					\includegraphics[scale=.87,valign=b]{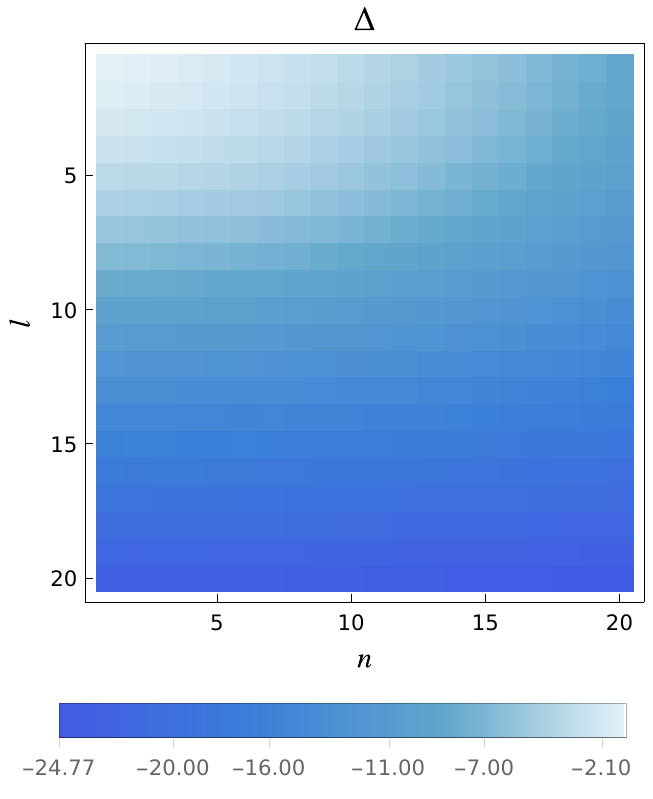}
					\end{tabular}
				}

				\end{tabular}
				\caption{Relative error $\Delta$ of number of excitations of non-relativistic approximation to full solution as a function of mode numbers $n$ and $l$.  Parameters are  $\varrho/L=0.5$,  $\Omega L =50$. (a, d)  $a L=0.00005$ (final velocity about $0.01$); 
					(b, e) $a L=0.005$ (final velocity about $0.1$);	(c, f) $a L=0.05$  (final velocity about $0.3$). The following field modes are within 2\%  of the detector's gap: $(l,n)=\{(1,16),(3,15),(4,14),(5,13),(6,11),(7,8),(7,9),(8,2),(8,3),(8,4)\}$.
					For the initially excited detector, the number expectation values are underestimated only for those modes with associated energies that are close to the detector's gap for low accelerations; for higher accelerations this will get shifted in $l$ direction mainly due to the Doppler shift. The higher energetic modes all have overestimated number expectations. In all cases, going to higher mode number $l$ or $n$ will result in a run-away relative difference from the exact solution even for low acceleration. For the initial ground-state setting of the detector, all number expectation values are overestimated with the relative difference being lowest for low energy modes, and again, even for low accelerations, the error is diverging from the exact solution by going to higher $l$ or $n$ modes.}
				\label{compNR1}
				\end{figure*}

				In the non-relativistic limit for $a \tau \ll 1$ the worldline \eqref{a world} approximates to
				\begin{align}
				z(\tau)&=\frac{a \tau^2}{2} + \mathcal{O}\left((a\tau)^2\right), ~t=\tau+\mathcal{O}\left((a \tau)^3\right),
				\end{align}
				and under this approximation, one can find an analytic solution to the integrals \eqref{number}   by noting that
				\begin{widetext}
				\begin{align}
				D_{\pm}(l, n) \coloneqq &\int_0^{T}\dd \tau e^{\ii \tau  (\omega\pm \Omega )} \sin(\frac{a \pi  n\tau ^2}{2 L})\nonumber \\
				=&\frac{(-1)^{\frac{1}{4}} \sqrt{L}}{2 \sqrt{2 a n}}\left\{  e^{\ii\frac{ L}{2 \pi  a n}\left(\omega\pm \Omega\right)^2} \left[\text{erf}\!\left(\frac{\frac{1+\ii}{2} (\pi  a n T-L (\omega\pm\Omega) )}{\sqrt{\pi a n L}}\right)+\text{erf}\!\left(\frac{\frac{1+\ii}{2} \sqrt{L} (\omega\pm\Omega )}{\sqrt{\pi a n}}\right)\right] \right.\nonumber\\
				&\quad\quad\quad\quad\quad\left.+\ii e^{-\ii\frac{L}{2 \pi  a n} (\omega\pm\Omega )^2} \left[\text{erf}\!\left(\frac{\frac{\ii-1}{2} (L(\omega\pm\Omega)+\pi  a n T )}{\sqrt{\pi a n L}}\right)-\text{erf}\!\left(\frac{\frac{\ii-1}{2} \sqrt{L} (\omega\pm\Omega )}{\sqrt{\pi a n }}\right)\right]\right\}.
				\end{align}
					\end{widetext}
				We can then write, recalling that only $m=0$ is non-vanishing,
				\begin{align}
				N_{l, n}\approx \lambda^2   |A_{0 l n}|^2  |D_{\pm}(l,  n)|^2. \label{NRdiag}
				\end{align}
			 The relative error $\Delta$  due to the non-relativistic approximation (denoted by superscript NR) for the expectation value of the number operator in the different field modes $N_{l,bn }$ is 
				\begin{align}
				\Delta(l, n)=1-\frac{N^{\text{NR}}_{l, n}}{N_{l, n}}.
				\end{align} 
				In Fig.~\ref{compNR1} we show the relative error for different proper accelerations $a$.  For the detector initially in excited and low numbers of $n$ and $l$, the error peaks at the most resonant modes.  In general, for low accelerations, we distinguish two kind of inaccuracies of the approximation. First, modes close in energy to the detector's gap from below will show an underestimation in the number expectation values.  Second, for the modes with larger values of $n$ and $l$, the non-relativistic approximation systematically overestimates  the number expectation values. Decreasing the proper acceleration reduces, as expected, the relative error in both cases. However, for the larger $n$ and $l$ set of modes, the approximation incurs in a run-away relative error from the exact values as we further increase the mode numbers $l$ or $n$. This means that for any fixed acceleration the approximation will fail to describe accurately the energy deposited for modes with $l$ and $n$ high enough.
				
				 If the detector starts out in the ground state, there is no such peaking but however the general overestimation of the number expectation value of the non-relativistic approximation can be seen. Towards larger values of $n$ and $l$ this overestimation increases, and again results in a diverging relative error for  unbounded mode numbers $l$ or $n$. Overall, the non-relativistic approximation is only reliable for low accelerations in the case of modes with low $n$ and $l$ values if the detector is initially in the ground state, and for modes which are close in energy to the detector's gap (but not resonant) for an initially excited detector.

				\section{1-dimensional approximation to long and thin cavities}\label{fibre}
				
Another common approximation that we see in the literature is to model an optical cavity through a 1+1D cavity instead of the more realistic 3+1D model. The question we address in this section is: can we perhaps approximate a very long and thin cavity (think of an  idealized, fully reflective optical fibre perhaps) by just a 1+1D cavity?

In more concrete words: we call the \textit{ idealized optical fibre limit} the limit of a very thin and very long cylindrical cavity. For $\varrho \ll L$, we want to see whether the model is effectively that of a massive scalar field in 1+1D. In that case the main contribution to the dynamics would then be dominated by the modes with $l=1$ (recall $l$ is the label of the radial modes). This is so because for any fixed $\Omega$, as we take the limit even the lowest energy radial mode will become far off-resonant with the detector gap, i.e. $\omega \gg \Omega$ as the  idealized optical fibre limit is taken.
				
				In particular, the energy spacing corresponding to modes with different quantum numbers $l$, coming from the radial modes, is significant.  
				This can be seen from \eqref{energ} as for low $n$ in the idealized optical fibre limit
				\begin{align}
				    \omega &\approx \frac{x_{0 l}}{\varrho} = \omega_0 \frac{x_{0 l}}{x_{0 1}},
				\end{align}
				where we have defined $\omega_0 \coloneqq x_{0 1} / \varrho$. Nonetheless, as we will see in the plots,   the first few $l$-modes greater than one will still be moderately excited by detector crossing.
				
\begin{figure*}
				\centering
								 Detector initially excited \\
				\begin{tabular}[c]{cc}
				\begin{tabular}[b]{c}
				\subfloat[ $\varrho/L=1/50$, $a L=1/20000$  ]{\label{fig:right}%
					\hspace*{-0.4cm}	\includegraphics[scale=.65,valign=b]{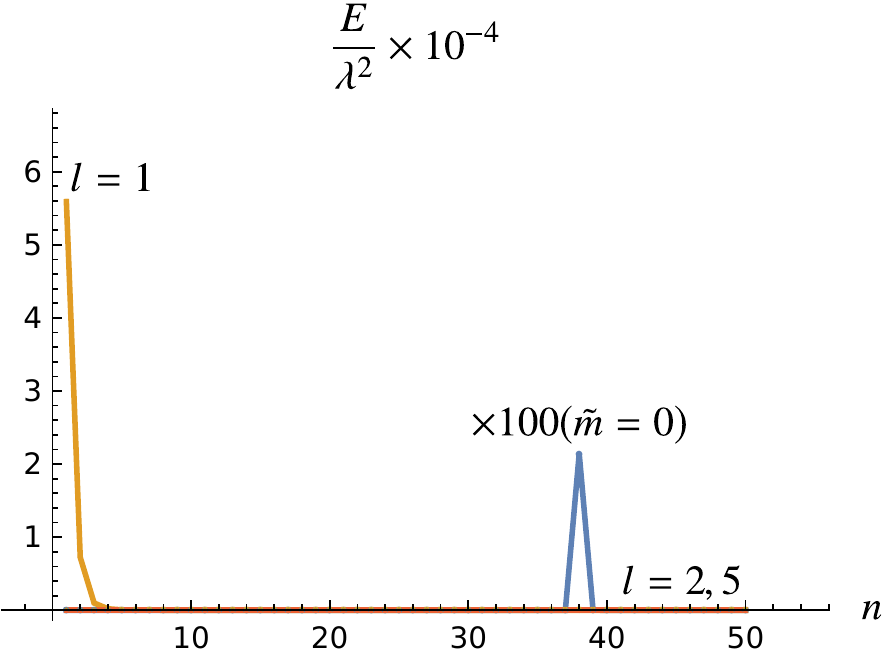}}
				
				\end{tabular}
				&
				\subfloat[ $\varrho/L=1/150$, $a L=1/20000$ ]{\label{fig:left}%

					\begin{tabular}[b]{c}
					\includegraphics[scale=.65,valign=b]{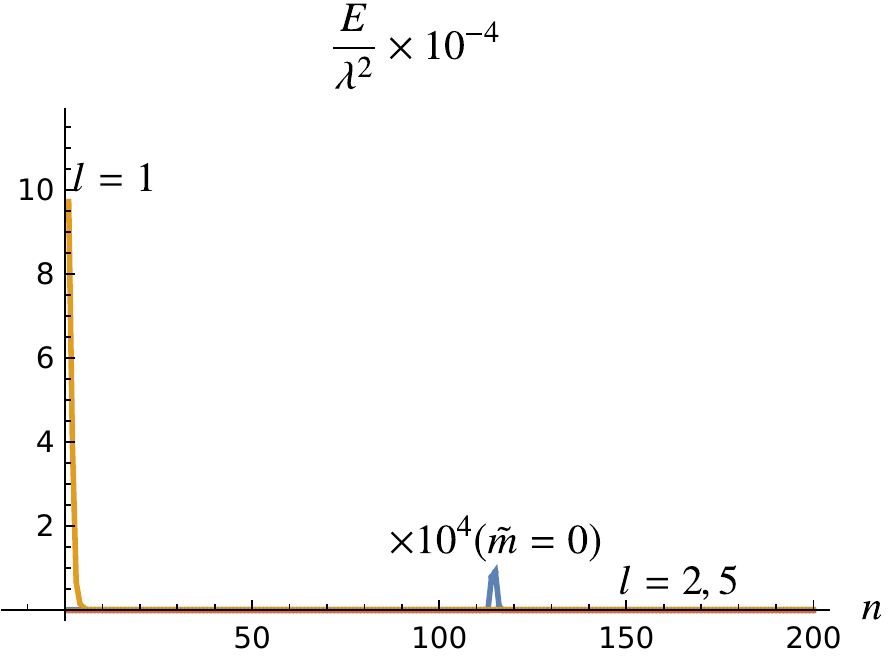}
					\end{tabular}
				}
				
				\subfloat[  $\varrho/L=1/50$, $a L=1/2 $ ]{\label{fig:left}%

					\begin{tabular}[b]{c}
					\includegraphics[scale=.65,valign=b]{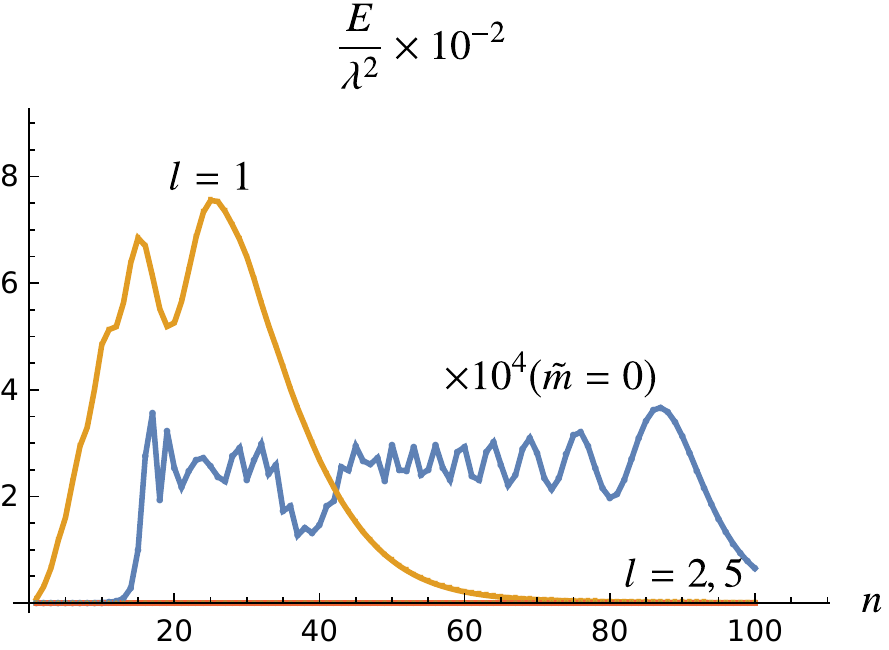}
					\end{tabular}
				}

				\end{tabular}
				 \centering
				 
				 \vspace{.5cm}
				 Detector initially in ground state \\
				\begin{tabular}[c]{cc}
				\begin{tabular}[b]{c}
				\subfloat[   $\varrho/L=1/50$, $a L=1/20000$   ]{\label{fig:right}%
					\hspace*{-0.4cm}	\includegraphics[scale=.65,valign=b]{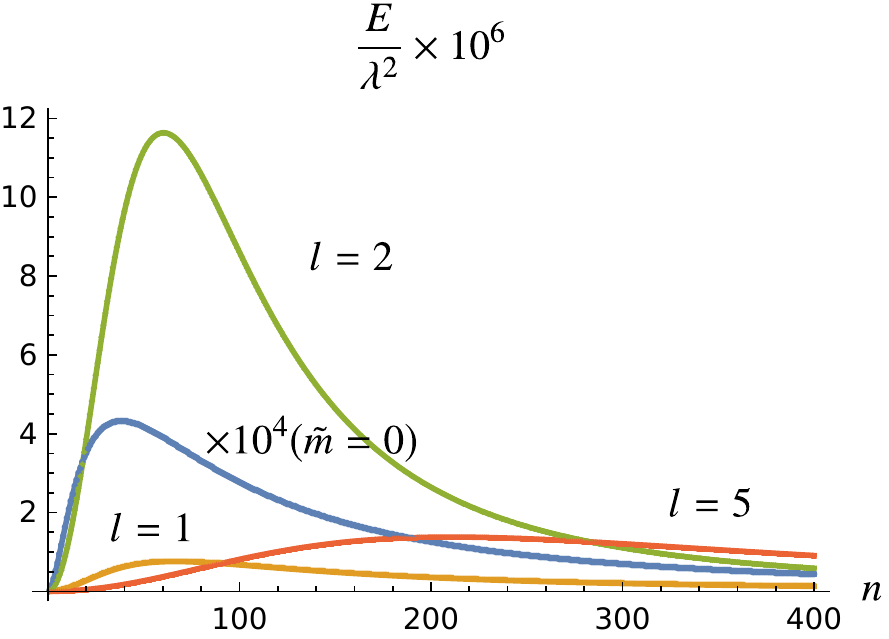}}
				
				\end{tabular}
				&
				\subfloat[$\varrho/L=1/150$, $a L=1/20000$ ]{\label{fig:left}%

					\begin{tabular}[b]{c}
					\includegraphics[scale=.65,valign=b]{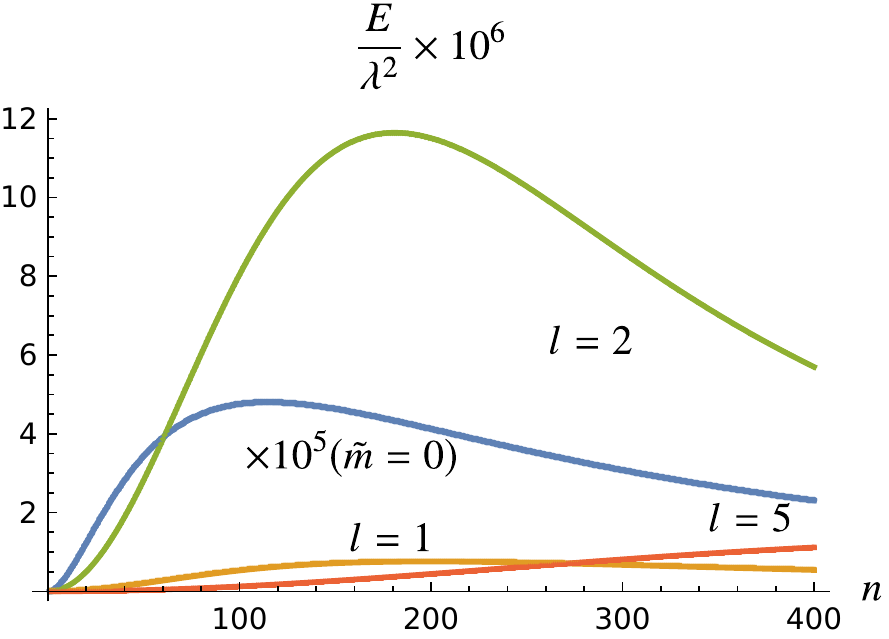}
					\end{tabular}
				}
				
				\subfloat[  $\varrho/L=1/50$, $a L=1/2 $ ]{\label{fig:left}%

					\begin{tabular}[b]{c}
					\includegraphics[scale=.65,valign=b]{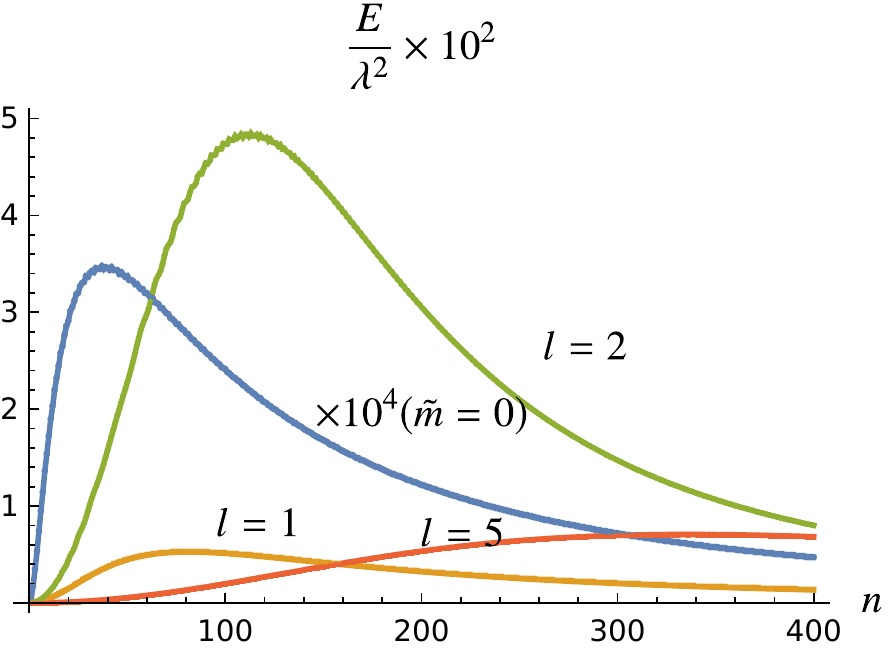}
					\end{tabular}
				}

				\end{tabular}
				\caption{Energy $E/ \lambda^2$ deposited in the field modes $n$ for different given fixed $l$ in the 3+1D model for different values of the radius $\varrho$ and acceleration $a$. Parameters are such that the detector's energy gap is most resonant with $(l,n)=(1,1)$ ($\Omega L =120.2$ in (a, c, d, f) and $\Omega L =360.7$ in (b, e)). The maximum depends on radial extension $\varrho$ and proper detector acceleration $a$. If the detector is initially in the ground state, the modes with $l=1$ are not the most dominant ones. As a comparison the deposited energy for a massless scalar field in the 1+1D model denoted by $\tilde{m}=0$ and rescaling for presentation purposes has been included. The behavior of the energy in the massless case deviates significantly (most prominently if the detector is initially excited), having the peak for a different $n$ than in the 3+1D case.}
				\label{nfixed}
				\end{figure*}

		As we will see below  in more detail (Fig.~\ref{nfixed} and discussion in the text), the best possible approximation for a 3+1D cavity with a  lower dimensional one is certainly not that of a 1+1D massless field. Instead, the transversal modes will introduce a term that would effectively act as a mass term in the equation of motion. Hence, we wish to compare to a scalar field with the effective mass $\tilde m$ in a cavity of length $L$ in 1+1D with Dirichlet boundary conditions. There the mode functions read 
			\begin{align}
			    \tilde u_n(z,t)= \frac{1}{\sqrt{\tilde \omega L }} e^{-\ii \tilde \omega t} \sin(\frac{n \pi}{L} z),
			\end{align}
			where $\tilde \omega=\sqrt{\tilde m^2 + (\frac{n \pi}{L})^2}$. Accordingly, we will choose $\tilde m=\tilde m(l)=\omega_0 x_{0 l}/x_{0 1}$. It follows that the mode functions of both models for a given $l$ (i.e., for a given mass $\tilde m(l)$),  when the detector in the  $3+1$D follows a longitudinal trajectory, are equal up to an ($l$-dependent) proportionality constant (cf. \eqref{modes}):
			\begin{align}
			    \tilde u_n(z,t)= \sqrt{\pi} J_1(x_{0 l}) \varrho  u_{0, l, n }(0, 0, z, t).\label{factor}
			\end{align}
		Notice that the two mode functions depend on different powers of $\varrho$. The effective 1+1D case inherits dependence on $\varrho$ through the mass $\tilde m$. For $\varrho \ll L$, $\tilde u_n(z,t) \propto \sqrt{\varrho}$ and $u_{0, l, n }(0, 0, z, t) \propto \varrho^{-1/2}$. On the other hand, as can be seen from \eqref{longit}, taking the limit $\varrho/L \rightarrow 0$ will increase the frequency of the oscillatory behavior resulting in enhanced cancellations of positive and negative contributions. Therefore, the deposited energy  in the field modes of the 3+1D as well as for the 1+1D will become smaller as the quantity $\varrho/L$ goes to 0. 
		 
			\begin{figure*}
				\centering
			
				 			 Detector initially excited \\
				 \begin{tabular}[c]{cc}
				 \begin{tabular}[b]{c}
				 \subfloat[ $\varrho/L=1/50$, $a L=1/20000$  ]{\label{fig:right}%
				 	\hspace*{-0.4cm}	\includegraphics[scale=.63,valign=b]{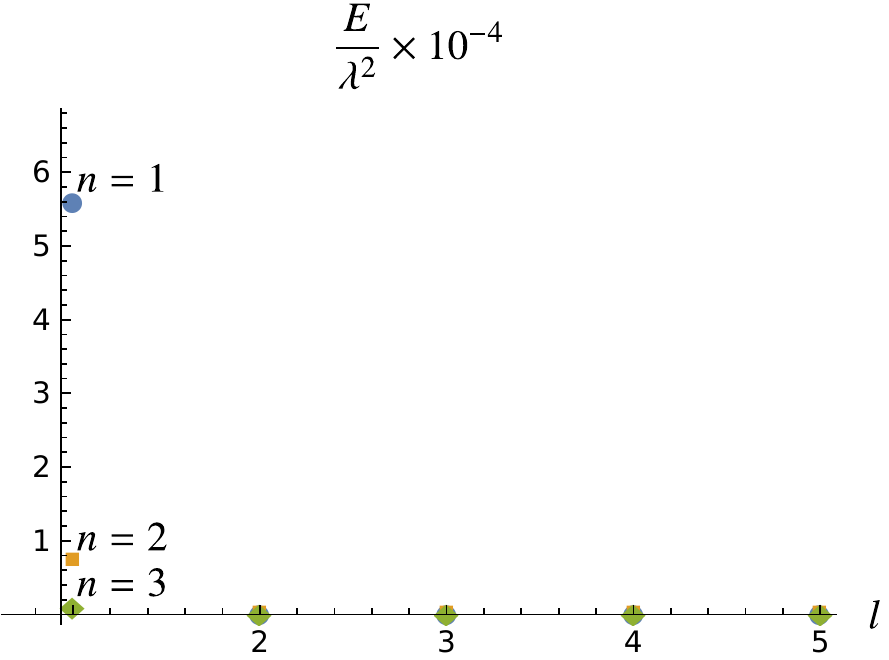}}
				
				 \end{tabular}
				 &
				 \subfloat[ $\varrho/L=1/150$, $a L=1/20000$ ]{\label{fig:left}%

				 	\begin{tabular}[b]{c}
				 	\includegraphics[scale=.63,valign=b]{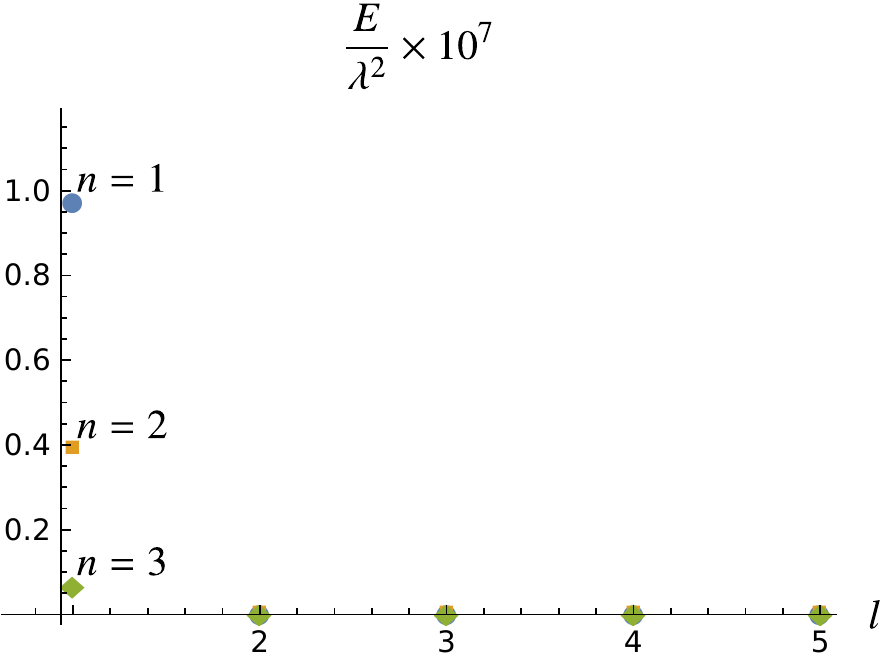}
				 	\end{tabular}
				 }
				
				 \subfloat[  $\varrho/L=1/50$, $a L=1/2 $ ]{\label{fig:left}%

				 	\begin{tabular}[b]{c}
				 	\includegraphics[scale=.63,valign=b]{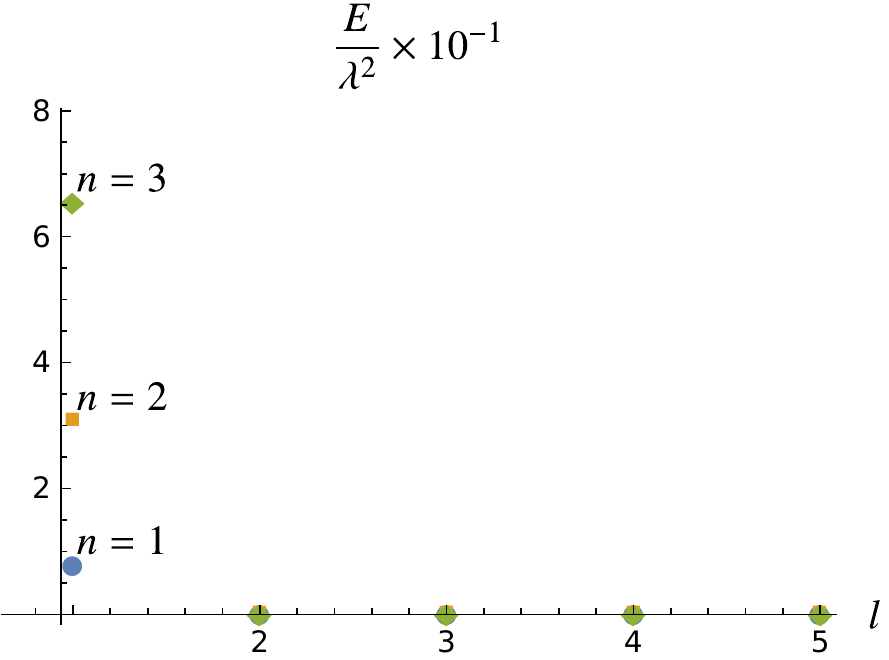}
				 	\end{tabular}
				 }

				 \end{tabular}
				  \centering
				 
				  \vspace{.5cm}
				  Detector initially in ground state \\
				 \begin{tabular}[c]{cc}
				 \begin{tabular}[b]{c}
				 \subfloat[   $\varrho/L=1/50$, $a L=1/20000$   ]{\label{fig:right}%
				 	\hspace*{-0.4cm}	\includegraphics[scale=.63,valign=b]{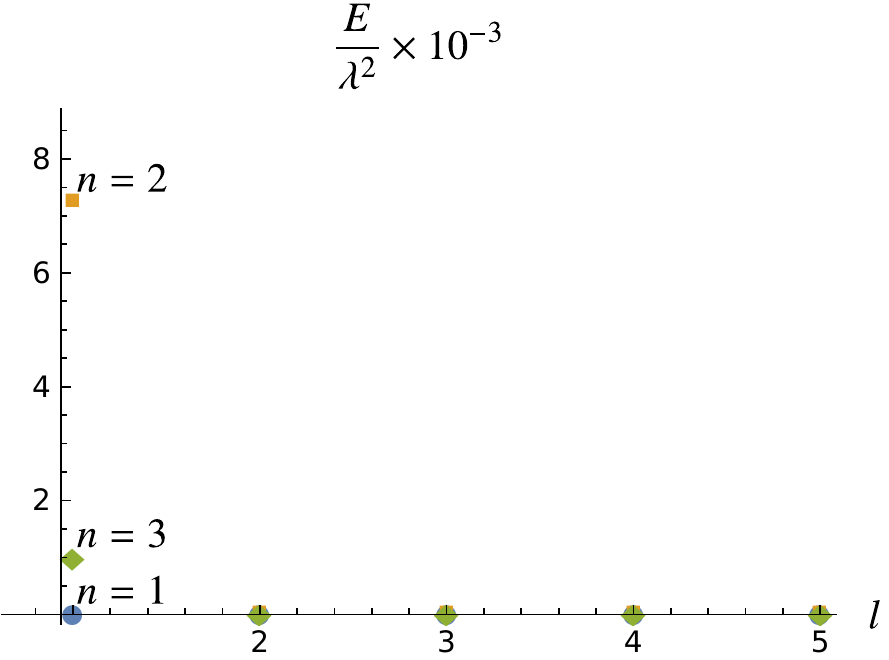}}
				
				 \end{tabular}
				 &
				 \subfloat[$\varrho/L=1/150$, $a L=1/20000$ ]{\label{fig:left}%

				 	\begin{tabular}[b]{c}
				 	\includegraphics[scale=.63,valign=b]{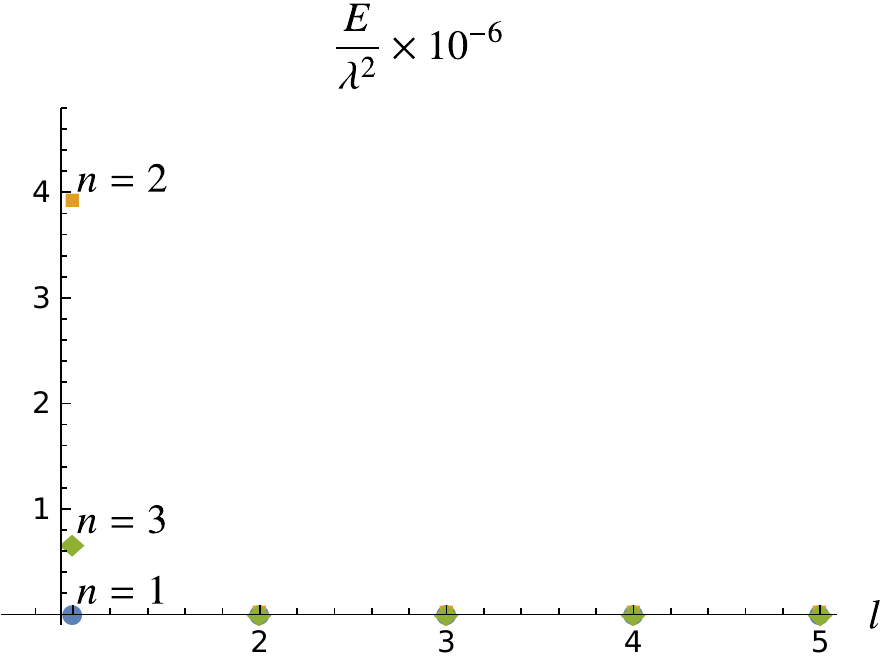}
				 	\end{tabular}
				 }
				
				 \subfloat[  $\varrho/L=1/50$, $a L=1/2 $ ]{\label{fig:left}%

				 	\begin{tabular}[b]{c}
				 	\includegraphics[scale=.63,valign=b]{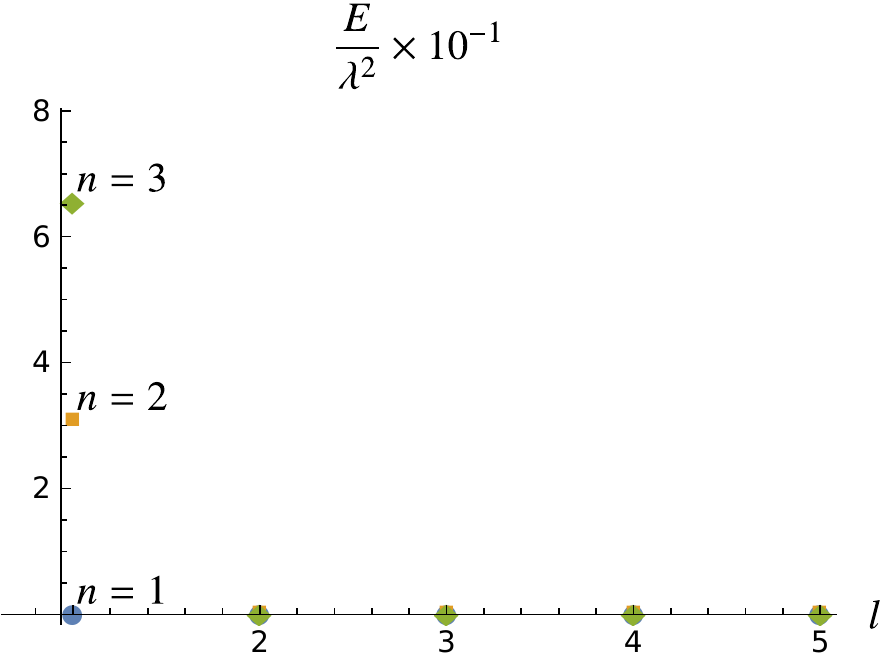}
				 	\end{tabular}
				 }

				\end{tabular}
				\caption{Energy $E/ \lambda^2$ deposited in the field modes $l$ for different given $n$ in the 3+1D model for different values of the radius $\varrho$ and acceleration $a$. Parameters are such that the detector's energy gap is most resonant with $(l,n)=(1,1)$ ($\Omega L =120.2$ in (a, c, d, f) and $\Omega L =360.7$ in (b, e)). The maximum is thus always located at $l=1$ for $\varrho/L \ll 1$. The modes with $n=1$ are only dominant for initially excited detectors for low accelerations.}
				\label{green}
				\end{figure*} 
		
		For fixed $l$, the behavior of the energy distribution in the field is shown in Fig.~\ref{nfixed}. If the detector starts out in the ground state, even though the energy gap $\Omega$ is resonant with the first field mode $(l,n)=(1,1)$, one can witness a peak in the deposited energy for modes with $l\neq 1$. The smaller $\varrho/L$ (in the case of ground-state detectors) or the larger the proper acceleration $a L$ of the detector, the more the peak for modes with fixed $l$ will occur for larger $n$. It also noticeable that for higher $l$, the peak will occur for larger values of $n$ and be less in magnitude for ground-state detectors. 
		Conversely, modes with $l>1$ are strongly suppressed if the detector is initially in the excited state. In Fig.~\ref{green}, the modes are shown for fixed $n$: The dominating contribution comes from $l=1$, and very quickly falls off for higher values of $l$. If the detector is initially excited for small accelerations, the mode with $n=1$ will be the most significant one. On the other hand, for ground-state detectors and larger accelerations, other modes will be more relevant. Also changing the radial extension or the acceleration mainly changes the overall magnitude, not where the dominating contribution is to be found for fixed $n$.
		
		For comparison, the energy in the field after detector crossing in the 1+1D model with a massless scalar field is shown as well in Fig.~\ref{nfixed}. An initially excited detector will cause that the energy is sharply localized at the resonant field mode, showing a Doppler shift for relativistic accelerations. If the detector starts out in the ground state, the energy peaks again at the resonance but falls off much slower. The deposited energy in the massless model has a peak that differs in magnitude and position from the 3+1D model, this is especially significant if the detector is initially excited. Therefore a massless field is not a good effective model for the  idealized optical fibre limit of the 3+1D case.
		
		Of course the natural effective model, as we argued, should be the massive 1+1D field. 
		We do not need to plot the energy distribution for the 1+1D case because, as we see from \eqref{factor}, the modes are equal up to a factor of $\sqrt{\pi} J_1(x_{0 l}) \varrho$, and therefore the energy deposited per mode in the 3+1D case will be equal to the 1+1D case up to the square of that factor. Namely using \eqref{number}:
		\begin{equation}
		    \tilde{E}^{(1+1)}_{l, n}= \pi J_1(x_{0 l})^2 \varrho^2 E^{(3+1)}_{l, n}.
		\end{equation}
		
		As we see in Fig.~\ref{nfixed}, modes with $l>1$ can be comparable or even larger in magnitude to modes with $l=1$ but for different values of $n$. Overall, for the same $l$ the modes in longitudinal direction, i.e. with quantum number $n$, can be modelled by a 1+1D massive case, where for every $l$ another mass is required in 1+1D.
		
		However, it is not true that the 1+1D massive model is always a good model for a thin cylindrical cavity. To see this let us to consider the excitation probability of the detector after leaving the cavity, and conclude whether it is the same in both models. If the detector in 1+1D starts out in the ground state, the probability $\tilde{ \mathcal{P}}^{g\to e}_l$, for a fixed mass $\tilde m$ of the scalar field given by $l$, of finding it excited when leaving the cavity has this form to leading order in perturbation series:
				\begin{align}
				\tilde{\mathcal{P}}^{g\to e}_l&= \sum_{\substack{n=1}}^{\infty} |\bra{(0, n), e} \hat U^{(1)} \ket{0, g}|^2\nonumber\\
				&= \lambda^2   \sum_{\substack{n=1}}^{\infty}  \left|\int_0^T \dd \tau e^{\ii \Omega \tau} \tilde u_n(z(\tau),t(\tau))^*   \right|^2, 
			\end{align}
			where the $l$ dependence is implicit in the $\tilde{\omega}$ which depends on the mass $\tilde{m}$ as a function of $l$.
			From \eqref{factor} one can compare it to the probability of excitation $\mathcal{P}^{g\to e}$ for the 3+1D model (cf. \eqref{excprob}):
			\begin{align}
			\mathcal{P}^{g\to e}=  \sum_{\substack{l=1}}^{\infty}	\mathcal{P}^{g\to e}_l= \sum_{\substack{l=1}}^{\infty} \frac{1}{\pi J_1(x_{0 l})^2 \varrho^2} \tilde{\mathcal{P}}^{g\to e}_l,
				\end{align}
				where $\mathcal{P}^{g\to e}_l$ without tilde is the collective contribution of all modes to the excitation probability for a given $l$:
				\begin{equation}
				   \mathcal{P}^{g\to e}_l= \lambda^2   \sum_{\substack{n=1}}^{\infty}  \left|\int_0^T \dd \tau e^{\ii \Omega \tau}  u_{0, l, n }(0, 0, z, t)^*   \right|^2.
				\end{equation}
				If the excitation probability for the detector in 3+1D has significant contributions only for $l=1$, then the 1+1D model is a good approximation in the  idealized optical fibre limit to describe the detector response. An estimator of the validity of the approximation could be the relative magnitude $F$ of the $l>1$ modes contribution to the probability with respect to the contribution that comes from the modes with $l=1$ which constitute the 1+1D approximation as discussed above.
				
				The estimator is difficult to evaluate, but we can easily derive a lower bound to it by evaluating the partial sum up to $N_l$ and $N_n$:
				\begin{align}
				F= \frac{\sum_{\substack{l=2}}^{\infty} 	\mathcal{P}^{g\to e}_l}{	\mathcal{P}^{g\to e}_1} \geq \frac{\sum_{\substack{l=2}}^{N_l} 	\mathcal{P}^{g\to e, N_n}_l}{	\mathcal{P}^{g\to e}_1},
				\end{align}
				where $\mathcal{P}^{g\to e, N_n}_l$ indicates that the sum in $n$ is truncated at $N_n$.
				For simplicity, we will consider a detector with small constant velocity $v$, which will allow to feasibly perform the sum numerically. The contribution from modes with a fixed $l$ to the excitation probability in the 3+1D model takes the following form (see \eqref{constv}):
				\begin{align}
				    \mathcal{P}^{g\to e}_l=  \sum_{\substack{n=1}}^{\infty} \frac{2 \lambda^2 \pi ( n  \bar{v})^2}{\omega L^3 (\varrho \gamma	J_1(x_{0 l}))^2 } \frac{1+(-1)^{n+1} \cos(\left(\omega +\frac{\Omega}{\gamma} \right) \frac{L}{v })}{  \left[\left(\omega+\frac{\Omega}{\gamma}\right)^2-\left(\frac{n \pi v}{L}\right)^2\right]^2}. 
				\end{align}
				\renewcommand{\arraystretch}{1.2}
				    \begin{table}[ht]
				        \centering
				        \begin{tabular}{|c|| c|c|c|c|c|c| }
				           \hline 
				            Parameters  & \multicolumn{6}{c|}{$\bar{v}=0.005$ , $\Omega L=20$} \tabularnewline \hline
				           $\varrho/L$ & $5\times 10^{-1}$ & $10^{-2}$&$10^{-3}$&$10^{-4}$&$10^{-5}$&$10^{-6}$\\ \hline 
				           $ F \geq$ &151.47 &4.9 &4.15 &4.11 &4.05 & 3.16 \\\hline
				           \hline 
				           Parameters  & \multicolumn{6}{c|}{$\bar{v}=0.005$ , $\Omega L$ resonant with $(l,n)=(1,1)$} \tabularnewline \hline 
				           $\varrho/L$ & $5\times 10^{-1}$ & $10^{-2}$&$10^{-3}$&$10^{-4}$&$10^{-5}$&$10^{-6}$\\ \hline 
				           $ F \geq$ &31.31&16.42&17&16.88&16.65&12.11\\\hline
				        \end{tabular}
				        \caption{Estimating the validity of the 1+1D model by considering the ratio of excitation probabilities $F$ as a function of $\varrho/L$ for constant non-relativistic velocity $\bar{v}=0.005$ in the 3+1D model, and both fixed detector's gap $\Omega L=20$ as well as varying detector's gap such that it is always most resonant with the first field mode. We have chosen as the cut-offs for the sums over $n$ and $l$ $N_n=10^8$ and $N_l=250$, respectively.
				        }
				        \label{constvv}
				    \end{table}
				The results are shown in Table \ref{constvv}. We find that $F$ does not seem to have the limit zero, and that therefore the dimensional reduction to 1+1D will not reproduce the results faithfully as the excitation probability for the 3+1D case has in fact non-negligible contributions coming from $l>1$. Furthermore, we show the validity of the single-mode approximation in 1+1D by evaluating the resonant contribution of a single field mode to the total excitation probability  for  scalar fields of different masses $m$ in Table~\ref{1+1 a vary}. There we included also the contribution of the resonant to the total probability to find the detector in its ground state if it was initially excited ($\mathcal{P}^{e\to g}$), with an analogous derivation as in Appendix~\ref{derive longit} for the 3+1D model. We have picked the detector's gap such that it is resonant with the first field mode since this will give us an upper bound to the resonant contribution to the total excitation probability. This can be seen in Table~\ref{1+1 O vary}: The ratio $\tilde{\mathcal{P}}_{\text{res}}^{g\to e}/ \tilde{\mathcal{P}}^{g\to e}$ falls off for larger values of the gap $\Omega$. In particular this is due to choosing only one mode of the field. For the case of the detector initially excited, we find that the single mode approximation holds well for non-relativistic trajectories, and that the size of the detector's gap minimally changes the validity of the single-mode approximation.
				Overall, the larger the mass of the scalar field, or the gap (in the case of the excitation probability) of the detector, and the more relativistic the trajectory, the worse the single-mode approximation will be. 
				
				 In the case of the excitation probability, even a more generous assumption of taking all the modes that are within 20\% of $\Omega$ as resonant contribution shows that a few-mode approximation will not be sufficient to reproduce the exact results (see Table~\ref{1+1 O vary, 20}). In that case, nonetheless, the ratio is largely insensitive to the gap $\Omega$, which implies that especially for larger $\Omega$ the contribution is more distributed among several modes.
			 Therefore, the single-mode approximation cannot reproduce the expected results for a detector that is initially in the ground state, that is even true for non-relativistic trajectories.

					\renewcommand{\arraystretch}{1.2}
 				    \begin{table*}[ht!]
 				        \centering
 				        \begin{tabular}{|c|| c|c|c|c|c|c| }
 				           \hline 
 				           Parameters  & \multicolumn{6}{c|}{$m L=0$, $\Omega L=3.14$ (resonant with $l=1$)} \tabularnewline \hline 
 				           $a L$&$5\times 10^{-5}$&$5\times 10^{-4}$&$5\times 10^{-3}$&$5\times 10^{-2}$&$5\times 10^{-1}$&200\\\hline
 				           $\tilde{\mathcal{P}}_{\text{res}}^{e\to g}/ \tilde{\mathcal{P}}^{e\to g} $&1.000&1.000&1.000&0.997&0.746&0.052 \\\hline
 				           $\tilde{\mathcal{P}}_{\text{res}}^{g\to e}/ \tilde{\mathcal{P}}^{g\to e} $&0.522& 0.523& 0.522& 0.529& 0.526& 0.051  \\\hline
 				            \hline 
 				           Parameters  & \multicolumn{6}{c|}{$m L=2.41$ , $\Omega L=3.95$ (resonant with $l=1$)} \tabularnewline \hline 
 				           $a L$&$5\times 10^{-5}$&$5\times 10^{-4}$&$5\times 10^{-3}$&$5\times 10^{-2}$&$5\times 10^{-1}$&200\\\hline
 				            $\tilde{\mathcal{P}}_{\text{res}}^{e\to g}/ \tilde{\mathcal{P}}^{e\to g} $&1.000&1.000&1.000&0.993&0.664&0.031 \\\hline
 				           $\tilde{\mathcal{P}}_{\text{res}}^{g\to e}/ \tilde{\mathcal{P}}^{g\to e} $&0.334& 0.334& 0.336& 0.327& 0.2626& 0.03 \\\hline
 				            \hline 
 				           Parameters  & \multicolumn{6}{c|}{$m L=4.81$ , $\Omega L=5.74$ (resonant with $l=1$)} \tabularnewline \hline 
 				           $a L$&$5\times 10^{-5}$&$5\times 10^{-4}$&$5\times 10^{-3}$&$5\times 10^{-2}$&$5\times 10^{-1}$&200\\\hline \hline
 				            $\tilde{\mathcal{P}}_{\text{res}}^{e\to g}/ \tilde{\mathcal{P}}^{e\to g} $&1.000&1.000&1.000&0.973&0.515&0.008 \\\hline
 				           $\tilde{\mathcal{P}}_{\text{res}}^{g\to e}/ \tilde{\mathcal{P}}^{g\to e} $&0.133& 0.133& 0.132& 0.127& 0.091& 0.008 \\\hline \hline
 				           Parameters  & \multicolumn{6}{c|}{$mL=48.1$ , $\Omega L=48.19$ (resonant with $l=1$)} \tabularnewline \hline 
 				           $a L$&$5\times 10^{-5}$&$5\times 10^{-4}$&$5\times 10^{-3}$&$5\times 10^{-2}$&$5\times 10^{-1}$&200\\\hline 
 				            $\tilde{\mathcal{P}}_{\text{res}}^{e\to g}/ \tilde{\mathcal{P}}^{e\to g} $&1.000&0.994&0.594&0.097&0.004&$1.5\times 10^{-5}$ \\\hline
 				           $\tilde{\mathcal{P}}_{\text{res}}^{g\to e}/ \tilde{\mathcal{P}}^{g\to e} $& $2.6\times 10^{-4}$& $2.6\times 10^{-4}$&$2.6\times 10^{-4}$  & $2.5\times 10^{-4}$& $1.8\times 10^{-4}$ & $1.5\times 10^{-5}$\\\hline
 				        \end{tabular}
 				        \caption{1+1D variation in $aL$: Ratio of the contribution of the resonant modes to the total  spontaneous emission probability $\tilde{\mathcal{P}}^{e\to g}$ and vacuum excitation probability $\tilde{\mathcal{P}}^{g\to e}$, respectively, with different values of $m L$, $\Omega L$,  and varying detector acceleration $aL$. The detector's gap is chosen such that the first field mode, i.e. $l=1$, is most resonant with the gap.
 				        }
 				        \label{1+1 a vary}
 				    \end{table*}

\begin{table*}
\parbox{.45\linewidth}{
\centering
\begin{tabular}{|c|| c|c|c| c|}
 				           \hline 
 				           Parameters  & \multicolumn{4}{c|}{$m L=0$, $a L=5\times 10^{-5}$} \tabularnewline \hline 
 				           $\Omega L$&3.14 ($l$=1)&10 ($l$=3)&50 ($l$=16)&100 ($l$=32)\\\hline
 				            $\tilde{\mathcal{P}}_{\text{res}}^{e\to g}/ \tilde{\mathcal{P}}^{e\to g} $&1.000&0.996&1.000&1.000 \\\hline
 				           $\tilde{\mathcal{P}}_{\text{res}}^{g\to e}/ \tilde{\mathcal{P}}^{g\to e} $&0.522&0.131&0.023&0.012 \\\hline
 				            \hline 
 				           Parameters  & \multicolumn{4}{c|}{$m L=2.41$, $a L=5\times 10^{-5}$} \tabularnewline \hline 
 				           $\Omega L$&3.95 ($l$=1)&10 ($l$=3)&50 ($l$=16)& 100 ($l$=32)\\\hline
 				            $\tilde{\mathcal{P}}_{\text{res}}^{e\to g}/ \tilde{\mathcal{P}}^{e\to g} $&1.000&1.000&1.000&1.000 \\\hline
 				           $\tilde{\mathcal{P}}_{\text{res}}^{g\to e}/ \tilde{\mathcal{P}}^{g\to e} $&0.334&0.137&0.024& 0.012\\\hline
 				            \hline
 				            Parameters  & \multicolumn{4}{c|}{$m L=4.81$, $a L=5\times 10^{-5}$} \tabularnewline \hline 
 				           $\Omega L$&5.74 ($l$=1)&10 ($l=$3)&50 ($l=$16)&100 ($l=$32)\\\hline
 				            $\tilde{\mathcal{P}}_{\text{res}}^{e\to g}/ \tilde{\mathcal{P}}^{e\to g} $&1.000&0.995&1.000&1.000 \\\hline
 				           $\tilde{\mathcal{P}}_{\text{res}}^{g\to e}/ \tilde{\mathcal{P}}^{g\to e} $&0.133&0.135&0.012&0.012 \\\hline
 				            \hline 
 				            Parameters  & \multicolumn{4}{c|}{$m L=48.09$, $a L=5\times 10^{-5}$} \tabularnewline \hline 
 				           $\Omega L$&$48.19$ ($l$=1)&49 ($l=3$)&70 ($l=16$)&111 ($l=32$)\\\hline
 				            $\tilde{\mathcal{P}}_{\text{res}}^{e\to g}/ \tilde{\mathcal{P}}^{e\to g} $&1.000&1.000&1.000&1.000\\\hline
 				           $\tilde{\mathcal{P}}_{\text{res}}^{g\to e}/ \tilde{\mathcal{P}}^{g\to e} $&$2.6\times 10^{-4}$&0.002&0.016&0.013 \\\hline
 				           
 				        \end{tabular}
\caption{1+1D variation in $\Omega L$: Ratio of the contribution of the resonant modes to the total  spontaneous emission probability $\tilde{\mathcal{P}}^{e\to g}$ and vacuum excitation probability $\tilde{\mathcal{P}}^{g\to e}$, respectively, with different values of $\tilde{m} L$,  and varying detector's gap  $\Omega L$ in the non-relativistic regime.  We have selected as resonant contribution only the one field mode closest in energy to $\Omega$, indicated in the brackets for the corresponding detector's gap.
 				        }
 				        \label{1+1 O vary}
}
\hfill
\parbox{.45\linewidth}{
\centering
 				        \begin{tabular}{|c|| c|c|c|c|c| }
				           \hline 
				           Parameters  & \multicolumn{4}{c|}{$m L=0$, $a L=5\times 10^{-5}$} \tabularnewline \hline 
 				           $\Omega L$&3.14&10&50&100\\\hline
 				           $\tilde{\mathcal{P}}_{\text{res}}^{g\to e}/ \tilde{\mathcal{P}}^{g\to e}$&0.522&0.131&0.166& 0.153\\\hline
 				           Parameters  & \multicolumn{4}{c|}{ $m L=2.41$, $a L=5\times 10^{-5}$} \tabularnewline \hline 
 				           $\Omega L$&3.95&10&50&100\\\hline
 				           $\tilde{\mathcal{P}}_{\text{res}}^{g\to e}/ \tilde{\mathcal{P}}^{g\to e}$&0.334&0.137&0.168&0.154 \\\hline
 				           Parameters  & \multicolumn{4}{c|}{$m L=4.81$, $a L=5\times 10^{-5}$} \tabularnewline \hline 
 				           $\Omega L$&5.74&10&50&100\\\hline
 				           $\tilde{\mathcal{P}}_{\text{res}}^{g\to e}/ \tilde{\mathcal{P}}^{g\to e} $&0.133&0.135&0.171&0.155\\\hline
 				           Parameters  & \multicolumn{4}{c|}{$m L=48.1$, $a L=5\times 10^{-5}$} \tabularnewline \hline 
 				           $\Omega L$&48.19&49&70&111\\\hline
 				           $\tilde{\mathcal{P}}_{\text{res}}^{g\to e}/ \tilde{\mathcal{P}}^{g\to e} $&0.07&0.069&0.187&0.2\\\hline
 				        \end{tabular}
\caption{1+1D variation in $\Omega L$ with few-mode approximation: Ratio of the resonant contribution to the vacuum excitation probability $\tilde{\mathcal{P}}^{g\to e}$ of the 1+1D model with $a L=5\times 10^{-5}$,  and varying $\Omega L$ for different masses in the non-relativistic regime.   Here those field modes that are in $20 \%$ difference from the detector's gap are taken to be the resonant modes.
 				        }
 				         \label{1+1 O vary, 20}
}
\end{table*}

			\section{Conclusion}
			We have studied the imprint of an accelerated detector crossing a cavity on the quantum field, as well as the transition probabilities of the detector after crossing the cavity.  In particular we looked at relativistic and non-relativistic regimes and found that a sharp localization in the energy distribution of the field can only be given for initially excited detectors with non-relativistic accelerations. However, even in these settings, we saw that assuming a single-mode or few-mode approximation will not always -- depending on the specific parameters of detector and cavity --  yield a satisfactory reproduction of the general results. Moreover, as soon as we enter the relativistic regime or have a detector initially in the ground state, a restriction of the relevant field modes to one or a few will even in principle fail to predict the correct results.
			
			We compared the results to the signature of the field if a detector crosses with constant velocity, and found that the distributions can be distinguished in order to extract the acceleration-induced influence on the field state after the detector crosses the cavity. Furthermore, we have shown that non-relativistic approximations on the trajectory of the detectors crossing optical cavities yield incorrect results for high-energy modes, even for non-relativistic regimes ($aL\ll 1$).
			
			Finally, we studied the case of a cavity where its length is much larger than its radius, i.e. that of an `optical fibre', and showed that neither a massive nor (above all) a massless quantum field in 1+1D can be reliably used to study the results of the \textit{ idealized optical fibre limit}.

			These results can become particularly relevant now that there are proposals to assess Unruh and Hawking effect related phenomena using atoms and optical cavities. The main conclusion of this paper is that some of the most common approximations made in quantum optics have to be questioned for any experiments involving relativistic effects. 
			
		\section{Acknowledgments}
The authors thank Jorma Louko for an insightful discussion. This project is partially supported by the NSERC Discovery program. E.M-M acknowledges funding from his Ontario Early Researcher Award.

				\appendix	
		\begin{widetext}

					\section{Solving for massless scalar field in cylindrical cavity}\label{derivation}
	First, we have to solve the massless Klein-Gordon equation for a cylindrical cavity with Dirichlet boundary conditions.
	We thus have a massless scalar field $\phi$ in a cavity of length $L$ and radius $\varrho$ such that in cylindrical coordinates
	\begin{align}
	\phi(r,\varphi,z,t)=0 ~\text{ for }~ z=0,~ z=L,~ r=\varrho. 
	\end{align}
	The corresponding Klein-Gordon equation has the form
	\begin{align}
	\Box \phi= \frac{1}{r} \pdv{ }{r}\left(r \pdv{\phi}{r} \right)+\frac{1}{r^2} \pdv[2]{\phi}{\varphi} + \pdv[2]{\phi}{z} -\pdv[2]{\phi}{t}=0, \label{KG}
	\end{align}
	and can be solved assuming the following separation ansatz: \mbox{$	\phi(r,\phi,z,t)= R(r)\Psi(\varphi) Z(z) T(t)$}. Hence \eqref{KG} yields
	\begin{align}
	\frac{1}{r R} \pdv{ }{r}\left(r \pdv{R}{r} \right)+\frac{1}{r^2 \Psi} \pdv[2]{\Psi}{\varphi} + \frac{1}{Z }\pdv[2]{Z}{z} = \frac{1}{T}\pdv[2]{T}{t}.
	\end{align}
	Since the right-hand side depends only on $t$ and the left-hand side is independent of $t$, both sides must be equal to a separation constant $-\omega^2$. This gives
	\begin{align}
	T(t)= e^{\pm \ii \omega t },
	\end{align}
	where we omit here (to be computed later) the normalization constant. Therefore, \eqref{KG} can be simplified and reads
	\begin{align}
	\frac{1}{r R} \pdv{ }{r}\left(r \pdv{R}{r} \right)+\frac{1}{r^2 \Psi} \pdv[2]{\Psi}{\varphi} + \frac{1}{Z }\pdv[2]{Z}{z}+\omega^2=0.
	\end{align}
	Following the same prescription for $Z$, we find using the separation constant $\alpha^2$
	\begin{align}
	\frac{1}{Z }\pdv[2]{Z}{z}&=-\omega^2 + \alpha^2 \newline
	\Rightarrow Z(z)=\exp(\pm \ii \sqrt{\omega^2-\alpha^2}z).
	\end{align}
	Employing the Dirichlet boundary conditions $Z(0)=Z(L)=0$, one arrives at
	\begin{align}
	Z(z)= \sin(\sqrt{\omega^2-\alpha^2} z)=\sin(\frac{n \pi}{L} z),
	\end{align}
	where $\sqrt{\omega^2-\alpha^2}=\frac{n \pi}{L}$ and $n\in \mathbb{Z}^+$. Accordingly,
	\begin{align}
	\frac{r}{ R} \pdv{ }{r}\left(r \pdv{R}{r} \right)+\alpha^2 r^2=-\frac{1}{ \Psi} \pdv[2]{\Psi}{\varphi}=m^2,
	\end{align}
	such that 
	\begin{align}
	\Psi(\varphi)= e^{\pm \ii m \varphi}.
	\end{align}
	In order to have a single-valued solution, i.e. $\Psi(\varphi+2\pi)=\Psi(\varphi)$, $m\in \mathbb{Z}^+$. Finally, for the radial part
	\begin{align}
	0&=	\frac{1}{r R} \pdv{ }{r}\left(r \pdv{R}{r} \right)-\frac{m^2}{r^2} +\alpha^2  \Rightarrow 0= \pdv[2]{R}{r} +\frac{1}{r} \pdv{R}{r} + \left(\alpha^2 -\frac{m^2}{r^2}\right) R.
	\end{align}
	Substituting $\alpha r=x$ yields the Bessel differential equation which can be solved by the Bessel functions $J_m(\alpha r)$ and $Y_m(\alpha r)$ of the first and second kind, respectively. Requiring regularity at $r=0$ and the boundary condition $R(\varrho)=0$, we find
	\begin{align}
	R(r)=J_m\left(\frac{x_{m l}}{\varrho} r\right),
	\end{align}
	where $\alpha=\frac{x_{m l}}{\varrho}$, and $x_{m l}$ is the $l$-th zero of the $m$th Bessel function $J_m$. This gives 
	\begin{align}
	\omega=\sqrt{\frac{x^2_{m l}}{\varrho^2}+ \frac{n^2 \pi^2}{L^2}}.
	\end{align} 
	Ultimately, and including a normalization factor $A_{m l n }$, the solution for the scalar field modes is
	\begin{align}
	u_{m n l}(r,\varphi,z,t)=A_{m  l n}  e^{\ii m \varphi}   e^{ -\ii \omega t } \sin(\frac{n \pi}{L} z) J_m\left(\frac{x_{m l}}{\varrho} r\right), 
	\end{align}
	such that the quantized scalar field takes the form
	\begin{align}
	\hat{\phi}(r,\varphi,z,t)= \sum_{\substack{m=0 \\ n,l=1}}^{\infty}\left(	u_{m l n } \hat{a}_{m l n } + 	u^*_{m l n } \hat{a}^{\dagger}_{m l n }\right),
	\end{align}
	where the creation and annihilation operators $ \hat{a}^{\dagger}_{m l n }$ and $\hat{a}_{m l n }$ have with $[\hat{a}_{m l n },\hat{a}^{\dagger}_{m' l' n'}]=\delta_{m m'}\delta_{l l'}\delta_{n n'}$ the usual commutation relations.
	The normalization factors can be found using the Klein-Gordon inner product:
	\begin{align}
	\left(u_{m l n }, u_{m' l' n'} \right)&= \ii \!\int \!\dd V \!\left(u_{m l n }^* \pdv{u_{m' l' n'}}{t} - u_{m' l' n'} \pdv{u^*_{m l n }}{t}\right) \nonumber\\
	&=2 A^*_{m l n } A_{m' l' n'} \omega \int_0^s \dd r r J_m\left(\frac{x_{m l}}{\varrho} r\right) J_{m'}\left(\frac{x_{m' l'}}{\varrho} r\right)  \int_0^L \dd z \sin(\frac{n \pi}{L})\sin(\frac{n' \pi}{L}) \int_0^{2\pi} \dd \varphi e^{\ii (m'-m) \varphi} \nonumber\\
	&=\frac{4 \pi L \varrho^2}{4 } |A_{m l n }|^2 \omega J_{m+1}(x_{m l})^2\delta_{m m'}\delta_{l l'}\delta_{n n'}
	\end{align}
	where $\dd V$ is a spacelike hypersurface, and we used the Sturm-Liouville orthogonality condition
	\begin{align}
	\int_0^s \dd r r J_m\left(\frac{x_{m l}}{\varrho} r\right) J_m\left(\frac{x_{m j}}{\varrho} r\right)= \frac{\varrho^2}{2} \delta_{l j} J_{m+1}(x_{m l})^2. \label{sturm}
	\end{align}
	Thus if we impose delta-normalization for the modes $\left(u_{m l n}, u_{m' l' n'} \right)=\delta_{m m'}\delta_{l l'}\delta_{n n'}$, then the normalization factor is obained as
	\begin{align}
	A_{m l n}= \frac{1}{\varrho \sqrt{L \pi \omega} J_{m+1}(x_{m l}) }.
	\end{align}
	\phantom{    }

				\section{Time-evolved field state}
	In this section we derive the state of the field after detector crossing. We start from \eqref{field rho} and introduce the parameter $\alpha$ which is either 1 or 0 when the detector is initially in the ground or excited state, respectively. As the correction to the initial field state takes the following form:
	\begin{equation}
	\hat \rho_\phi-\ket 0\!\!\bra{0}  = \tr_{\text{d}}\left( \hat U^{(1)} \hat \rho_0 \hat U^{(1) \dagger}\right) + \left(\tr_{\text{d}}\left( \hat U^{(2)} \hat \rho_0\right) + \text{H.c.}\right) 
	\end{equation}
	 we find
		\begin{align}
		\hat U^{(1)} \hat \rho_0 \hat U^{(1) \dagger}&=\lambda^2 \int_0^T \dd \tau \int_0^T \dd \tau' \hat{\phi}(\bm x(\tau),t(\tau))\ket 0\!\!\bra{0} \hat{\phi}(\bm x(\tau'),t(\tau')) \hat\mu(\tau) (\delta_{\alpha 1} \ket g\!\!\bra g +\delta_{\alpha 0} \ket e\!\!\bra e) \hat\mu(\tau')\nonumber\\
		&=\lambda^2 \!\int_0^T \!\dd \tau \!\int_0^T \!\dd \tau' \!\left( \delta_{\alpha 1} e^{\ii \Omega(\tau-\tau')} \ket e\!\!\bra e +\delta_{\alpha 0}e^{-\ii \Omega(\tau-\tau')} \ket g\!\!\bra g  \right) \!\!\sum_{\substack{m=0 \\m'=0}}^{\infty} \sum_{\substack{n,l=1 \\ n',l'=1}}^{\infty}\! \! u^*_{m l n}(\tau) u_{m' l' n'}(\tau') \ket{(m l n)}\!\!\bra{(m' l' n')}.
		\end{align}
		Thus after tracing over the detector's degrees of freedom
		\begin{align} \tr_{\text{d}}\left(\hat U^{(1)} \hat \rho_0 \hat U^{(1) \dagger}\right)&=\lambda^2 \int_0^T \dd \tau \int_0^T \dd \tau'e^{\pm\ii \Omega(\tau-\tau')} \sum_{\substack{m=0 \\m'=0}}^{\infty} \sum_{\substack{n,l=1 \\ n',l'=1}}^{\infty} u^*_{m l n}(\tau) u_{m' l' n'}(\tau') \ket{(m l n)}\!\!\bra{(m' l' n')} \label{diag1},
		\end{align}
		where here, and in the following, the top sign of either $\pm$ or $\mp$ is associated with the initial ground state of the detector, and the bottom sign with its excited state.
		For the remaining correction term:
		\begin{align}
		\hat U^{(2)} \hat \rho_0&=-\lambda^2 \int_0^T \dd \tau \int_0^\tau \dd \tau' \hat{\phi}(\bm x(\tau),t(\tau)) \hat{\phi}(\bm x(\tau'),t(\tau')) \ket 0\!\!\bra{0}  \mu(\tau) \mu(\tau') (\delta_{\alpha 1} \ket g\!\!\bra g +\delta_{\alpha 0} \ket e\!\!\bra e) \nonumber\\
		&= -\lambda^2 \int_0^T \dd \tau \int_0^\tau \dd \tau' (\delta_{\alpha 1}  e^{-\ii \Omega(\tau-\tau')} \ket g\!\!\bra g +\delta_{\alpha 0}  e^{\ii \Omega(\tau-\tau')} \ket e\!\!\bra e) \nonumber\\
		&\quad\times\sum_{\substack{m=0 \\m'=0}}^{\infty} \sum_{\substack{n,l=1 \\ n',l'=1}}^{\infty} \left[u^*_{m l n}(\tau) u^*_{m' l' n'}(\tau') \ket{(m l n), (m' l' n')}\!\!\bra{0} + \delta_{m m'}  \delta_{l l'}\delta_{n n'} u_{m l n}(\tau) u^*_{m l n}(\tau') \ket 0 \!\!\bra{0} \right].
		\end{align}
		Then after tracing out it yields
		\begin{align}
		\tr_{\text{d}}\left( \hat U^{(2)} \hat \rho_0\right)=-\lambda^2 \int_0^T \dd \tau \int_0^\tau \dd \tau'e^{\mp\ii \Omega(\tau-\tau')} \sum_{\substack{m=0 \\m'=0}}^{\infty} \sum_{\substack{n,l=1 \\ n',l'=1}}^{\infty}\left[u^*_{m l n}(\tau) u^*_{m' l' n'}(\tau') \ket{(m l n),(m' l' n')}\!\!\bra{0} \right.\nonumber\\
		\left. + \delta_{m m'}  \delta_{l l'}\delta_{n n'} u_{m l n}(\tau) u^*_{m l n}(\tau')\ket{0}\!\!\bra{0}\right].
		\end{align}
		
		\subsection{Particularizing to longitudinal motion}\label{derive longit}
		We consider the case that the detector crosses the cavity longitudinally such that the field modes are as in \eqref{longit}. To study the number of excitations and corresponding energy deposited in each field mode after a single run we only need to be concerned with the diagonal elements $ \ket{(m l n)}\!\!\bra{(m l n)}$.
	 To that end, \eqref{diag1} can be separated in diagonal and off-diagonal terms, where the former takes the form
		\begin{align}
		\left.\tr_{\text{d}}\left( \hat U^{(1)} \hat \rho_0 \hat U^{(1) \dagger}\right)\right|_{\text{diag}}=\lambda^2 &\sum_{\substack{n,l=1}}^{\infty}      |A_{0 l n}|^2  \left|\int_0^T \dd \tau e^{\pm \ii \Omega\tau} e^{ \ii \frac{\omega}{a}\sinh(a \tau) } \sin(\frac{n \pi}{a L}\left(\cosh(a \tau)-1\right))\right|^2 \ket{(0 l n)}\!\!\bra{(0 l n)} . \label{diag}
		\end{align}

	Therefore the number expectation value in modes $n$, $l$ reads
	\begin{align}
	    N_{l, n}=\lambda^2 &  |A_{0 l n}|^2  \left|\int_0^T \dd \tau e^{\pm \ii \Omega\tau} e^{ \ii \frac{\omega}{a}\sinh(a \tau) } \sin(\frac{n \pi}{a L}\left(\cosh(a \tau)-1\right))\right|^2 , 
	    \end{align}
	where $l, n >0$.
	It is not necessary to study the contribution coming from $	\tr_{\text{d}}\left( \hat U^{(2)} \hat \rho_0\right)$ as the only diagonal term is $\ket{0}\!\!\bra 0$, which can be found by making use of the vanishing trace of the corrections to density operators in perturbation theory at every order respectively.

	The probability $\mathcal{P}^{g\to e}$ of finding the detector that was initially in its ground state excited after crossing the cavity is, to leading order, given by
				\begin{align}
				\mathcal{P}^{g\to e}&= \sum_{\substack{n,l=1}}^{\infty} |\bra{(0, n, m), e} \hat U^{(1)} \ket{0, e}|^2=\bra{e} 	\tr_{\phi}\left( \hat U^{(1)} \hat \rho_0 \hat U^{(1) \dagger}\right) \ket{e}= \lambda^2   \sum_{\substack{n,l=1}}^{\infty}  \left|\int_0^T \dd \tau e^{\ii \Omega \tau}  u_{0ln}^*(\tau)   \right|^2 = \sum_{\substack{n,l=1}}^{\infty} N_{l,n}. \label{excprob}
			\end{align}	
		Similarly, we find the probability $\mathcal{P}^{e\to g}$ of an initially excited detector to be in its ground state after leaving the cavity to be
		\begin{align}
				\mathcal{P}^{e\to g}&= \sum_{\substack{n,l=1}}^{\infty} |\bra{(0, n, m), g} \hat U^{(1)} \ket{0, g}|^2=\bra{g} 	\tr_{\phi}\left( \hat U^{(1)} \hat \rho_0 \hat U^{(1) \dagger}\right) \ket{g}
				= \lambda^2   \sum_{\substack{n,l=1}}^{\infty}  \left|\int_0^T \dd \tau e^{-\ii \Omega \tau}  u^*_{0ln}(\tau)  \right|^2= \sum_{\substack{n,l=1}}^{\infty} N_{l,n}. 
		\end{align}
		Therefore for each corresponding initial detector setting  and to the leading order in perturbation series that we are using, the total expectation number of excitations in the field equals either of the detector probabilities. Hence, the resonant mode contribution to the total excitations in the field equals the ratio of the resonant mode contribution to the total vacuum excitation or spontaneous emission probability, depending on the initial detector state (suppressing the superscripts):
		\begin{align}
		    \frac{N_{\text{res}}}{\sum_{n,l=1}^\infty N_{l,n}} =\frac{\mathcal{P}_\text{res}}{\mathcal{P}}.
		\end{align}
		
		\end{widetext}
		\section{Parameter space}\label{parameterspace}
			
\begin{figure*}[!h]
				\centering
				\textbf{Large radial extension $\varrho \gg L$}\par\medskip
				Detector initially in excited state ~~~~~~~~~~~~~~~~~~~~~~~~~~~Detector initially in ground state \\
				\begin{tabular}[c]{cccc}
				\subfloat[  $a L=0.00005$ ]{\label{fig:right}%
				\includegraphics[scale=0.8,valign=t]{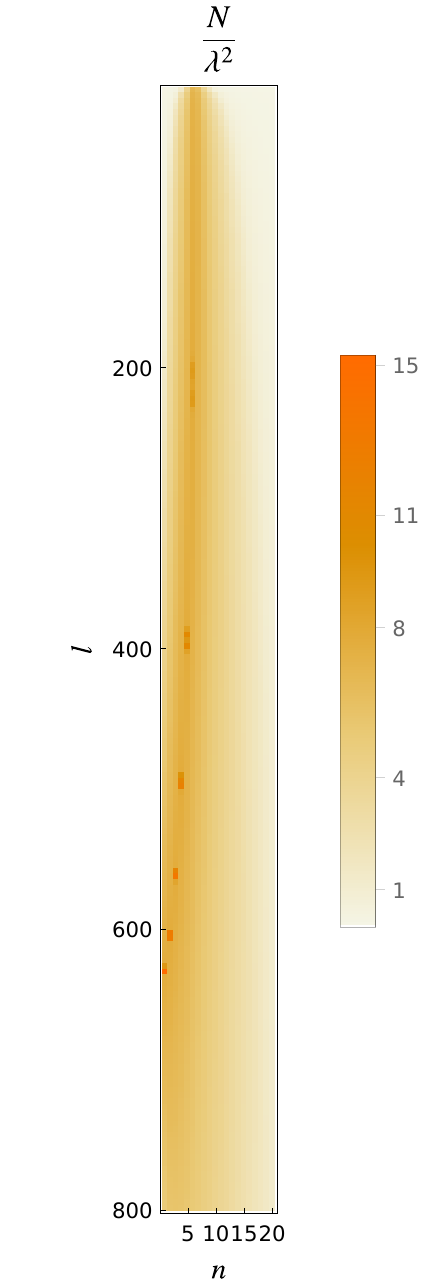}}
			& \subfloat[ $a L=0.5$ ]{\label{fig:left}%
			\includegraphics[scale=0.8,valign=t]{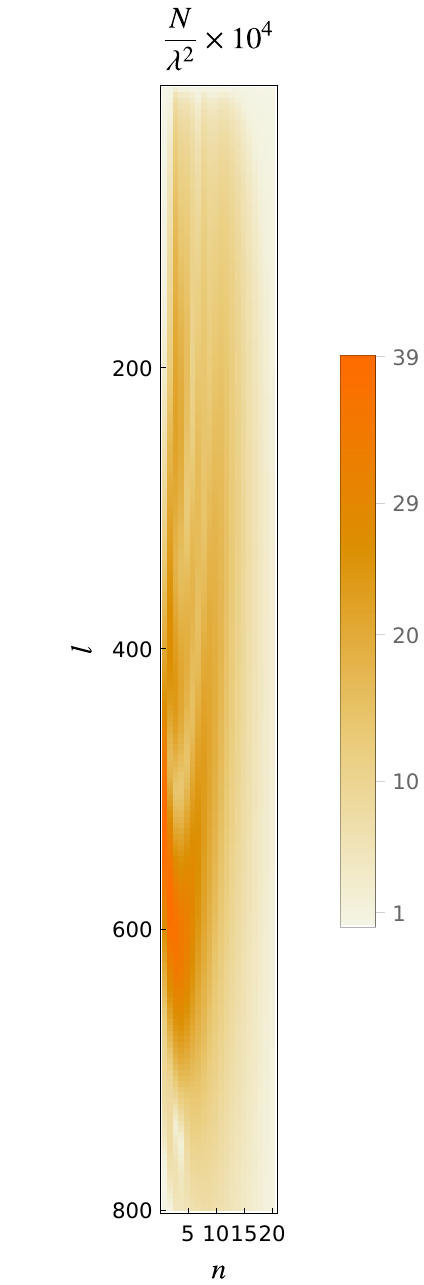}	}
			&\subfloat[  $a L=0.00005$ ]{\label{fig:right}%
				\includegraphics[scale=0.8,valign=t]{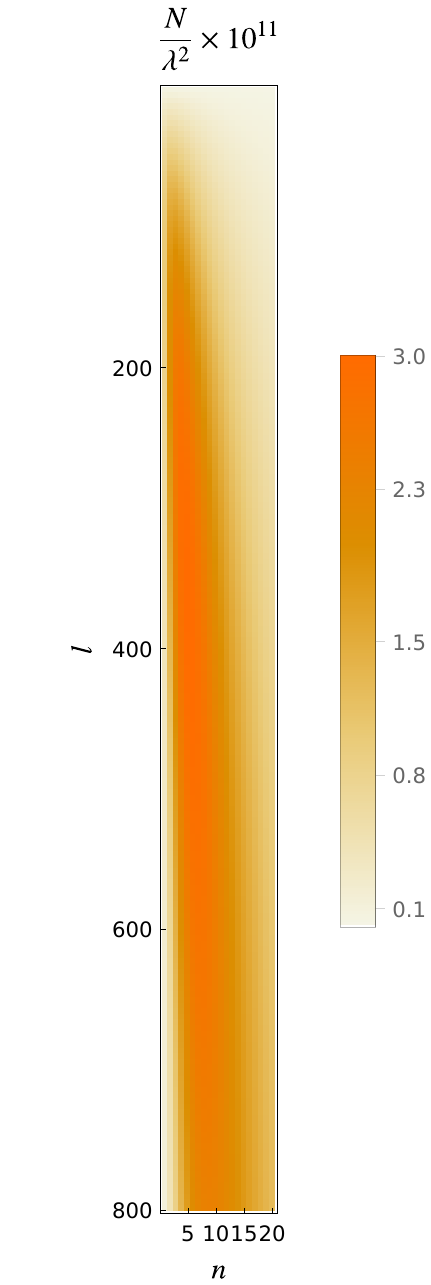}}
			& \subfloat[ $a L=0.5$ ]{\label{fig:left}%
			\includegraphics[scale=0.8,valign=t]{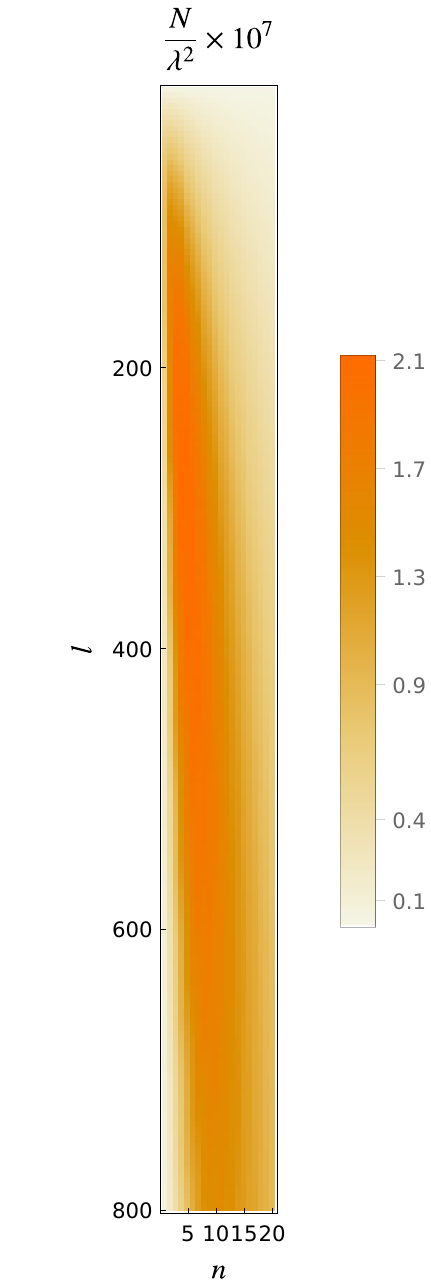}	}
				\end{tabular}
				\caption{Number expectation value $N$ as a function of mode numbers $n$ and $l$ for an exemplary $\varrho/L \gg 1$ setting. The parameters are $\varrho/L=100$, $\Omega L=20$ such that the detector's energy gap is most resonant with $(l, n)=\{(212,6),(213,6),(214,6),(394,5),(395,5),(495,4),(496,4),(562,3),(604,2),(605,2),(629,1)\}$ (assuming at most a $0.1$\%-difference in energy from the detector's gap).}
				\label{larges}
				\end{figure*}

\begin{figure*}[!h]
				\centering
				Detector initially in excited state \\
			\begin{tabular}[c]{cc}
				\begin{tabular}[b]{c}
				\subfloat[   $a/ \Omega=1.7 \times 10^{-6}$   ]{\label{fig:right}%
					\hspace*{-0.4cm}	\includegraphics[scale=.87,valign=b]{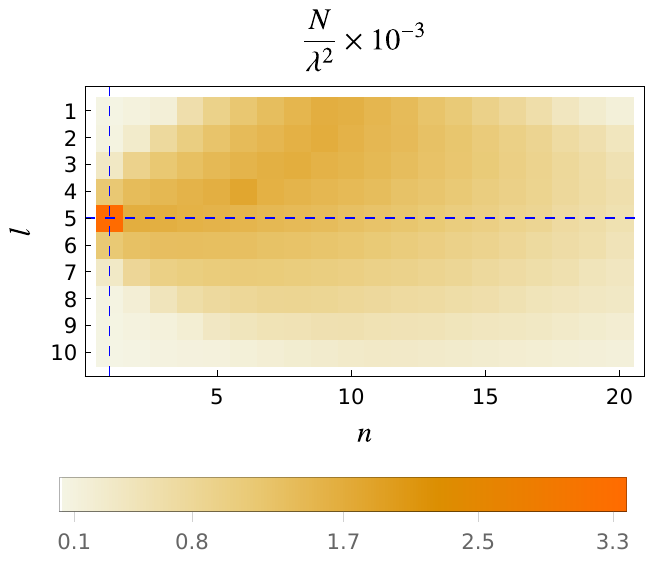}}
				
				\end{tabular}
				&
				\subfloat[ $a/ \Omega=10^{-4}$]{\label{fig:left}%

					\begin{tabular}[b]{c}
					\includegraphics[scale=.87,valign=b]{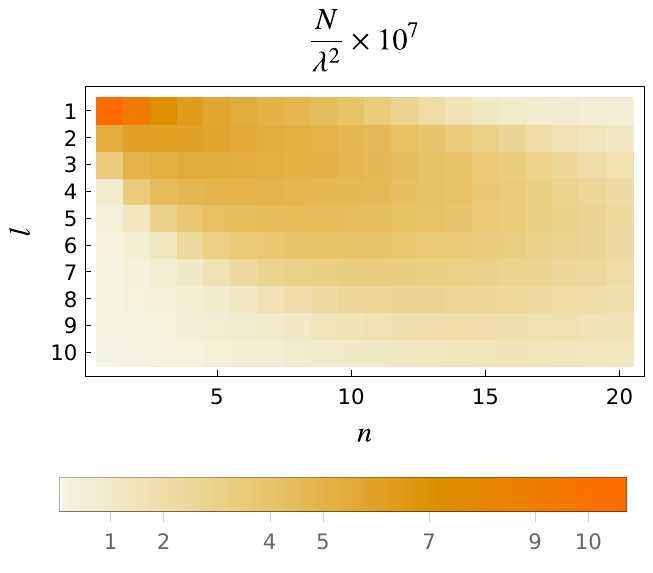}
					\end{tabular}
				}
				
				\subfloat[ $a/ \Omega=5 \times 10^{-2}$ ]{\label{fig:left}%

					\begin{tabular}[b]{c}
					\includegraphics[scale=.87,valign=b]{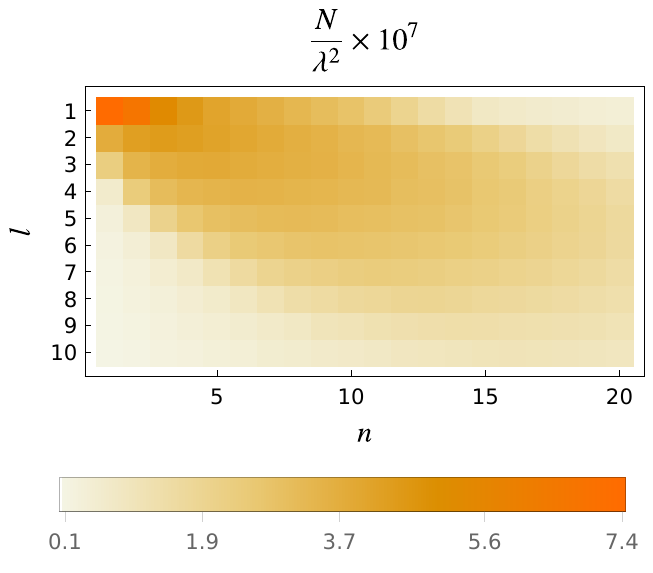}
					\end{tabular}
				}

				\end{tabular}
				 \centering
				 
				 \vspace{.5cm}
				 Detector initially in ground state \\
				\begin{tabular}[c]{cc}
				\begin{tabular}[b]{c}
				\subfloat[  $a/ \Omega=1.7 \times 10^{-6}$   ]{\label{fig:right}%
					\hspace*{-0.4cm}	\includegraphics[scale=.87,valign=b]{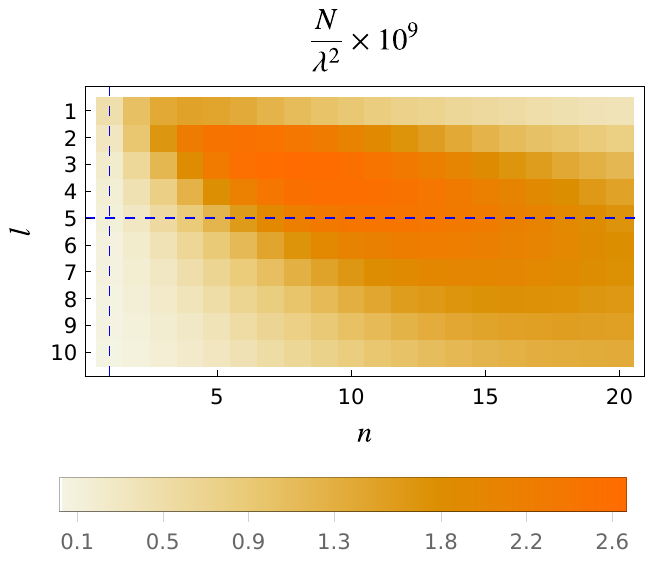}}
				
				\end{tabular}
				&
				\subfloat[$a/ \Omega=10^{-4}$ ]{\label{fig:left}%

					\begin{tabular}[b]{c}
					\includegraphics[scale=.87,valign=b]{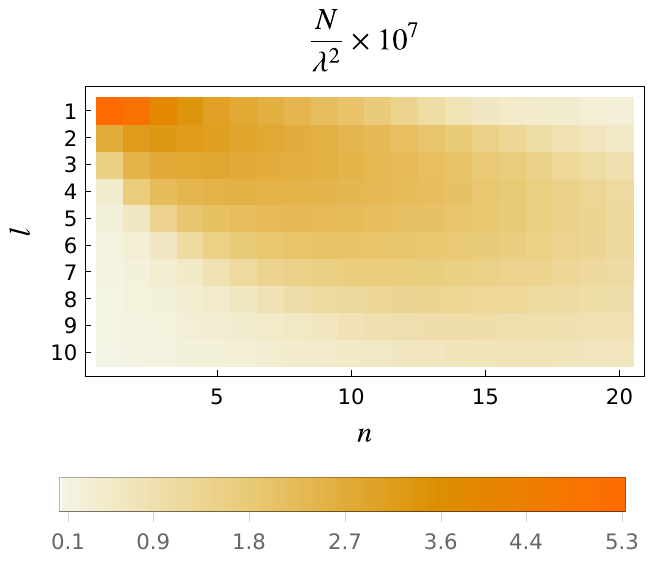}
					\end{tabular}
				}
				
				\subfloat[  $a/ \Omega=5 \times 10^{-2}$ ]{\label{fig:left}%

					\begin{tabular}[b]{c}
					\includegraphics[scale=.87,valign=b]{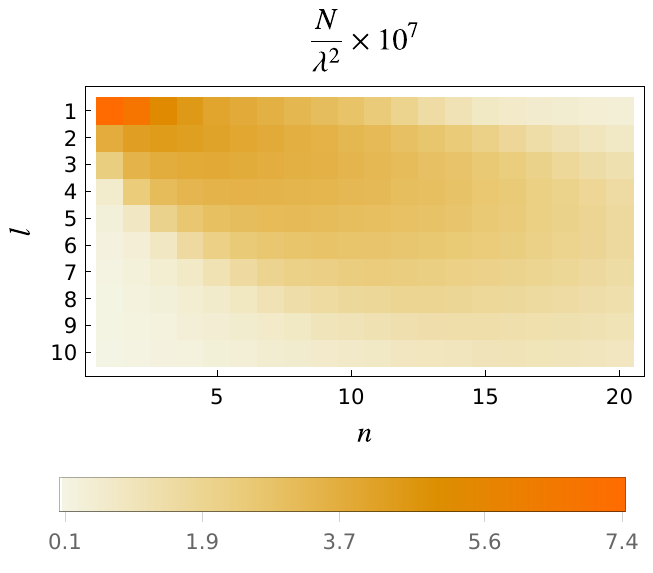}
					\end{tabular}
				}

				\end{tabular}
				\caption{Number expectation value $N$ as a function of mode numbers $n$ and $l$. Parameters are $\varrho/L=0.5$, $aL=0.00005$, and varying $a/\Omega$ such that the detector's gap is (a, d)  most resonant with $(l, n)=(1, 5)$ (intersection of dashed line); (b, c, e, f)  off-resonant with any field mode.}
				\label{parameter}
				\end{figure*}

\begin{figure*}[!h]
				\centering
				Detector initially in excited state \\
			\begin{tabular}[c]{cc}
				\begin{tabular}[b]{c}
				\subfloat[  $\Omega/\omega_0=0.8$   ]{\label{fig:right}%
					\hspace*{-0.4cm}	\includegraphics[scale=.87,valign=b]{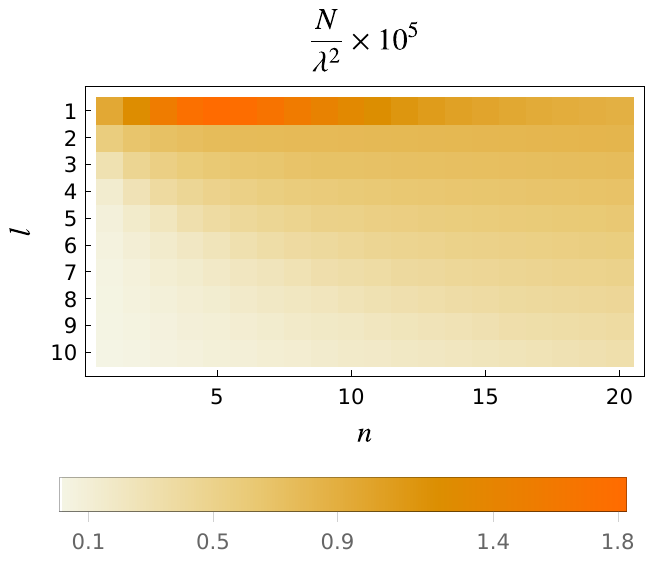}}
				
				\end{tabular}
				&
				\subfloat[ $\Omega/\omega_0=1.7$ ]{\label{fig:left}%

					\begin{tabular}[b]{c}
					\includegraphics[scale=.87,valign=b]{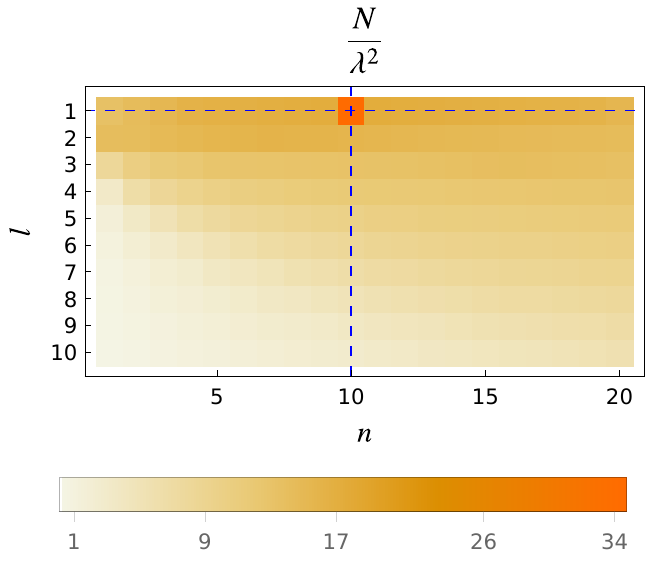}
					\end{tabular}
				}
				
				\subfloat[ $\Omega/\omega_0=8.3$ ]{\label{fig:left}%

					\begin{tabular}[b]{c}
					\includegraphics[scale=.87,valign=b]{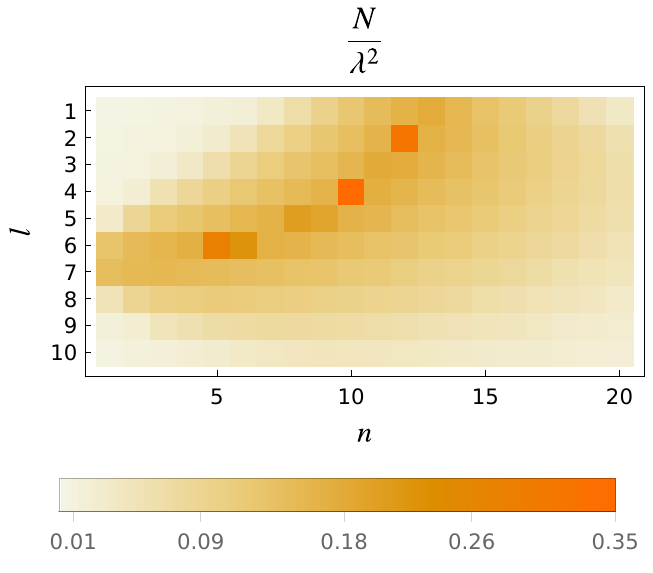}
					\end{tabular}
				}

				\end{tabular}
				 \centering
				 
				 \vspace{.5cm}
				 Detector initially in ground state \\
				\begin{tabular}[c]{cc}
				\begin{tabular}[b]{c}
				\subfloat[ $\Omega/\omega_0=0.8$   ]{\label{fig:right}%
					\hspace*{-0.4cm}	\includegraphics[scale=.87,valign=b]{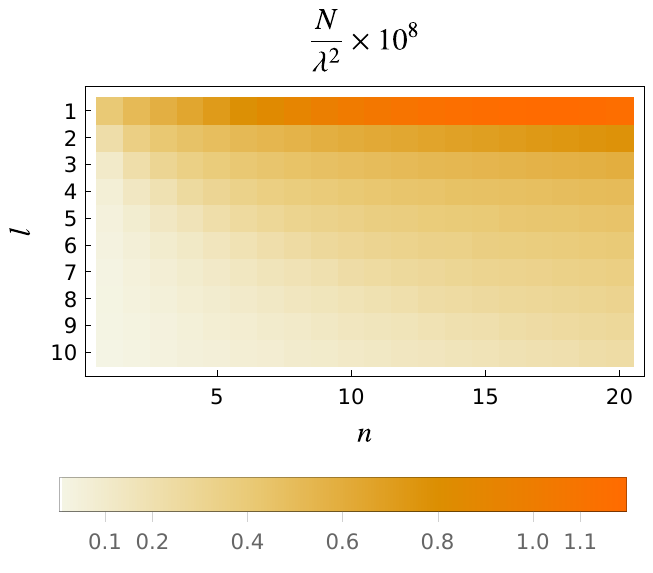}}
				
				\end{tabular}
				&
				\subfloat[$\Omega/\omega_0=1.7$  ]{\label{fig:left}%

					\begin{tabular}[b]{c}
					\includegraphics[scale=.87,valign=b]{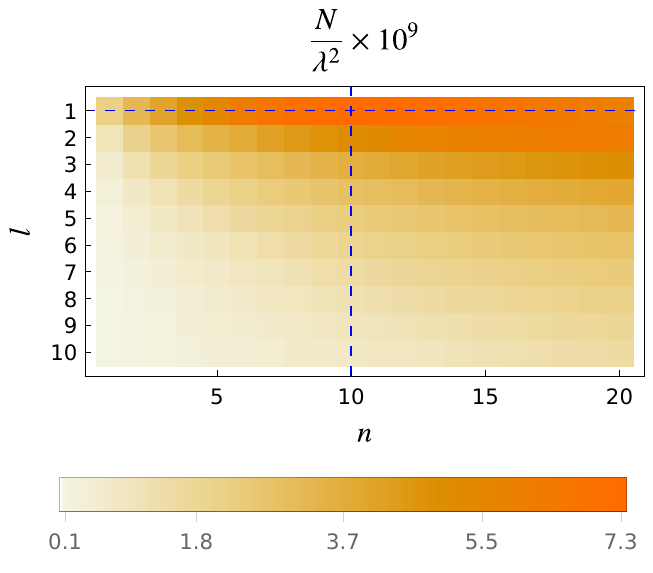}
					\end{tabular}
				}
				
				\subfloat[  $\Omega/\omega_0=8.3$ ]{\label{fig:left}%

					\begin{tabular}[b]{c}
					\includegraphics[scale=.87,valign=b]{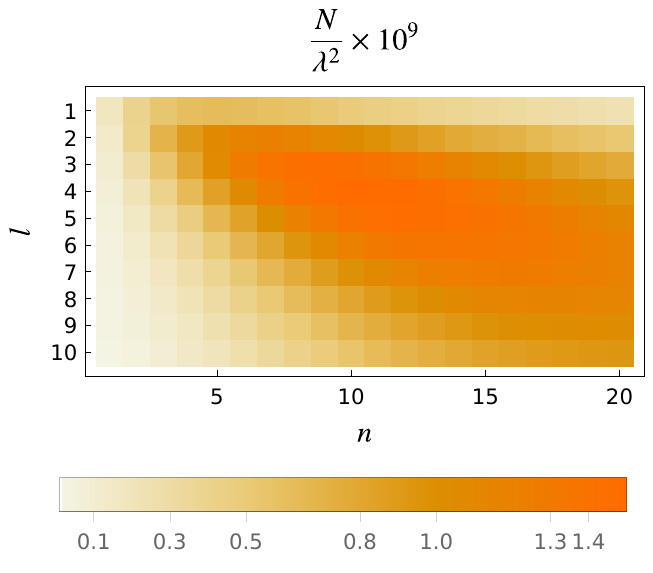}
					\end{tabular}
				}

				\end{tabular}
				\caption{Number expectation value $N$ as a function of mode numbers $n$ and $l$. Parameters are $a L=0.00005$, $\Omega L=40$ and varying $\Omega/\omega_0$. The detector's gap $\Omega$ is (a, d) off-resonant with any field mode; (b, e) most resonant with $(l, n)=(1, 10)$ (intersection of dashed line); (c, f) most resonant with $(l,n)=(2, 12), (4, 10), (6, 5)$.}
				\label{parameter2}
				\end{figure*}

			In this appendix we look at different dimensionless parameters, in contrast to the previously used measure for relativistic behavior $a L$, and investigate the influence on the expected number of excitations in the field after the detector's crossing. In particular, the following parameters will be studied: $a/\Omega$, $\Omega L$, $\Omega/ \omega_0$, and $L/\rho$. The results are collected in Figures~\ref{larges}, \ref{parameter}, and \ref{parameter2}. It is recognizable that, as the `lack of resonance' intuition dictates, if the detector's energy gap is small compared to the energy of any field mode, the excitations are centered around $(l, n)=(1, 1)$ (the closest-to-zero energy mode), spread over many modes, and there is not much qualitative difference between excited and ground state. 
			
			When the gap is comparable with a cavity mode energy we observe the same phenomenology as in Fig.~\ref{compes}: The excited detector releases energy in the modes close to resonance  and changing the detector's gap will change the location of the resonance, and the Doppler shift for it. Conversely, given a detector gap resonant with a field mode, a detector flying into the cavity in its ground state does not  effect any narrowly localized field excitations. 
			
			If we analyze the dependence on the cavity width,  $\varrho/L$, taking the opposite limit to Sec.~\ref{fibre}, i.e.  $\varrho \gg L$, we see (Fig.~\ref{larges}) that excitations are mainly localized at modes of constant $n$ corresponding to resonance with $\Omega$ distributed in a very spread manner among a variety of $l$ modes  if the detector is initially excited.
			Conversely, a initially ground-state detector excites the field in a big spread of modes, the same as what we saw before.
			It may be emphasized that, as expected, the number expectation values are larger for an excited detector as compared to one in the ground state, usually by several orders of magnitude. Moreover, increasing the acceleration increases the number expectation values for both detector settings.

						In the following, we will study once more the ratio of the resonant contribution to the transition probabilities. This will yield additional insight into the validity of the single-mode approximation. We choose as resonant modes those which are within 2\% of the detector's gap. In case no mode fulfills the criterion, we choose the one closest in energy. On the other hand, in certain regimes there will be more modes added as resonant than is actually justified from the energy distribution point of view. However, this will only strengthen our argument. In Table~\ref{ratiopara} we present the ratio for when several of the dimensionless parameters, i.e. $a/\Omega$, $\Omega L$, $\Omega/ \omega_0$ and $L/\rho$, are either very small or very large. It can be seen that, for an initially accelerated detector, with non-relativistic trajectories the single-mode approximation may be sufficient, depending on the specific parameters. However, as soon as the acceleration is increased the approximation fails inevitably. If the detector is initially in the ground state, the single-mode approximation will not even for non-relativistic trajectories reproduce the exact results.
			
							\renewcommand{\arraystretch}{1.2}
				    \begin{table*}[ht]
				        \centering
				        \begin{tabular}{|c|| c|c|c|c|c|c| c|c|c|}
				           \hline 
				         Limits  & $\Omega=0$&$L\ll \varrho, 1/\Omega$&$\varrho\ll L, 1/\Omega$&\multicolumn{3}{c|}{$L \approx \varrho\approx 1/\Omega$}&$L\gg \varrho, 1/\Omega$&\multicolumn{2}{c|}{$\varrho\gg L, 1/\Omega$}\\ \hline
				         Resonant modes  & off-resonant  & off-resonant &off-resonant& 10 modes&10 modes&10 modes&146 modes&680 modes&680 modes\\ \hline
				           $a L$ & $5\times 10^{-5}$  & $5\times 10^{-11}$ &$5\times 10^{-5}$&$5\times 10^{-5}$&$5\times 10^{-4}$&$5\times 10^{-3}$&$5\times 10^{-2}$&$5\times 10^{-5}$&$5\times 10^{-4}$\\ \hline 
				           $a/\Omega$ & $\infty$  & $2.5\times 10^{-6}$ & $2.5\times 10^{-6}$& $10^{-6}$&$10^{-5}$&$10^{-4}$&$5\times 10^{-6}$&$5\times 10^{-6}$&$5\times 10^{-5}$\\ \hline 
				           $\Omega L$ & 0 &$2\times 10^{-5}$ &20&50&50&50&$10^{4}$&10&10\\ \hline 
				           $\Omega/\omega_0$ & 0  &4.16 &$8.3\times 10^{-3}$&10.4&10.4&10.4&2.08&4158.3&4158.3\\ \hline 
				           $\varrho/L$ & 0.5 &$5\times 10^5$ &$10^{-3}$&0.5&0.5&0.5&$5\times 10^{-4}$&$10^3$&$10^3$\\ \hline 
				           $ \mathcal{P}^{e\to g}_\text{res}/\mathcal{P}^{e\to g} \leq$ &$6.5\times 10^{-2}$ &$3.1\times 10^{-5}$ &$2.9\times 10^{-9}$ & 0.9999&0.978 &0.53&$7\times 10^{-2}$&0.999&0.88\\
				           \hline
				           $\mathcal{P}^{g\to e}_\text{res}/\mathcal{P}^{g\to e}\leq$&$6.9\times 10^{-2}$ & $3.1\times 10^{-5}$&$2.8\times 10^{-9}$&$6.5\times 10^{-5}$&$6.5\times 10^{-5}$&$6.5\times 10^{-5}$&$5.4\times 10^{-3}$&$1.7\times 10^{-2}$&$1.7\times 10^{-2}$ \\\hline
				        \end{tabular}
				        \caption{Estimating an upper bound to the ratio of the resonant contribution to the full transition probabilities. We have chosen those modes for the resonant contribution which differ in energy from $\Omega$ by at most $2\%$.
				        In case there is no mode resonant with the detector's energy gap (first 3 cases), we have chosen $(l,n)=(1,1)$ as closest in energy to the detector's gap. As the cut-offs for the sums over $n$ and $l$ we have $10^4$ and $200$, respectively. Note however that for the 8th and 9th case we used $10^4$ and 4000 as cut-offs for the sums over $n$ and $l$, respectively. 
				        }
				        \label{ratiopara}
				    \end{table*}
				    
				    
\clearpage
									\bibliography{myref}

				\end{document}